\documentclass[a4paper,11pt]{article}
\pdfoutput=1 

\usepackage{jcappub} 

\usepackage[T1]{fontenc} 
\usepackage{lipsum}
\usepackage{graphicx}
\usepackage{subcaption}
\usepackage[sort&compress]{natbib}
\usepackage{bm}
\graphicspath{{paper_plots/}}
\usepackage{tikz}
\usepackage{simpler-wick}
\usepackage{subcaption}
\usepackage[margin=1.0in]{geometry}
\usetikzlibrary{shapes.geometric}

\newcommand{\ABL}[1]{{\textcolor{green}{\textbf{ABL}: #1}}}

\newcommand{\clbbdelunbias}{C_l^{BB,\text{del}, \text{unbiased}}}
\newcommand{\meas}{\frac{d^2\bm{l}}{(2\pi)^2}}
\newcommand{\measp}{\frac{d^2\bm{l}'}{(2\pi)^2}}

\newcommand{\measppp}{\frac{d^2\bm{l}'''}{(2\pi)^2}}
\newcommand{\measpppp}{\frac{d^2\bm{l}\prim{4}}{(2\pi)^2}}

\newcommand{\obs}{\text{obs}}
\newcommand{\lat}{\text{LAT}}
\newcommand{\fid}{\mathrm{fid}}
\newcommand{\sat}{\text{SAT}}
\newcommand{\res}{\text{res}}
\newcommand{\temp}{\text{temp}}

\newcommand{\tot}{\text{tot}}
\newcommand{\We}{\mathcal{W}^{E}}
\newcommand{\Welat}{\We}
\newcommand{\Wp}{\mathcal{W}^{\phi}}
\newcommand{\bl}{\bm{l}}
\newcommand\prim[1]{^{%
		\ifcase#1 \or\prime\or\prime\prime\or\prime\prime\prime\else\mathrm{\romannumeral #1}\fi}}

\def \corrOI{\begin{tikzpicture}[scale=.4, transform shape, baseline=0pt]
\tikzstyle{every node} = [circle]
\node[fill=red!50,star] (a) at (120:2cm) {B};
\node[fill=red!50] (b) at (180:2cm) {E};
\node[fill=red!50] (c) at (240:2cm) {E};
\node[fill=red!50] (d) at (300:2cm) {E};
\node[fill=red!50] (e) at (360:2cm) {E};
\node[fill=red!50,star] (f) at (420:2cm) {B};

\foreach \from/\to in {a/b, b/c, c/d, d/e, e/f, f/a}
\draw [-] (\from) -- (\to);
\end{tikzpicture}}

\def \corrOII{\begin{tikzpicture}[scale=.4, transform shape, baseline=0pt]
	\tikzstyle{every node} = [circle]
	
	\node[fill=red!50,star] (a) at (120:2cm) {B};
	\node[fill=red!50] (b) at (180:2cm) {E};
	\node[fill=red!50] (c) at (240:2cm) {E};
	\node[fill=red!50] (d) at (300:2cm) {E};
	\node[fill=red!50] (e) at (360:2cm) {E};
	\node[fill=red!50,star] (f) at (420:2cm) {B};
	\path (b) edge[out=93,in=87, looseness=2.1] node {} (e);
	\foreach \from/\to in {a/c, c/d, d/f, f/a}
	\draw [-] (\from) -- (\to);
	\end{tikzpicture}}
\def \corrOIII{\begin{tikzpicture}[scale=.4, transform shape, baseline=0pt]
	\tikzstyle{every node} = [circle]
	
	\node[fill=red!50,star] (a) at (120:2cm) {B};
	\node[fill=red!50] (b) at (180:2cm) {E};
	\node[fill=red!50] (c) at (240:2cm) {E};
	\node[fill=red!50] (d) at (300:2cm) {E};
	\node[fill=red!50] (e) at (360:2cm) {E};
	\node[fill=red!50,star] (f) at (420:2cm) {B};
	
	\foreach \from/\to in {a/f, b/e, c/d}
	\draw [-] (\from) -- (\to);
	\end{tikzpicture}}
\def \corrIII{	\begin{tikzpicture}[scale=.4, transform shape, baseline=0pt]
	\tikzstyle{every node} = [circle]
	
	\node[fill=red!50, star] (a) at (120:2cm) {B};
	\node[fill=red!50] (b) at (180:2cm) {E};
	\node[fill=red!50] (c) at (240:2cm) {E};
	\node[fill=red!50] (d) at (300:2cm) {E};
	\node[fill=red!50] (e) at (360:2cm) {E};
	\node[fill=red!50, star] (f) at (420:2cm) {B};
	
	\foreach \from/\to in {a/b, b/e, c/d, e/f, f/a}
	\draw [-] (\from) -- (\to);
	\end{tikzpicture}}
\def \corrI{\begin{tikzpicture}[scale=.4, transform shape, baseline=0pt]
	\tikzstyle{every node} = [circle] 
	\node[fill=red!50, star] (a) at (120:2cm) {B};
	\node[fill=red!50] (b) at (180:2cm) {E};
	\node[fill=red!50] (c) at (240:2cm) {E};
	\node[fill=red!50] (d) at (300:2cm) {E};
	\node[fill=red!50] (e) at (360:2cm) {E};
	\node[fill=red!50, star] (f) at (420:2cm) {B};
	
	\foreach \from/\to in {e/b, b/c, c/d, d/e, f/a}
	\draw [-] (\from) -- (\to);
\end{tikzpicture}}
\def \corrII{\begin{tikzpicture}[scale=.4, transform shape, baseline=0pt]
	\tikzstyle{every node} = [circle]
	
	\node[fill=red!50, star] (a) at (120:2cm) {B};
	\node[fill=red!50] (b) at (180:2cm) {E};
	\node[fill=red!50] (c) at (240:2cm) {E};
	\node[fill=red!50] (d) at (300:2cm) {E};
	\node[fill=red!50] (e) at (360:2cm) {E};
	\node[fill=red!50, star] (f) at (420:2cm) {B};
	
	\foreach \from/\to in {a/f, b/c, d/e}
	\draw [-] (\from) -- (\to);
	\end{tikzpicture}}
\def \corrIV{\begin{tikzpicture}[scale=.4, transform shape, baseline=0pt]
	\tikzstyle{every node} = [circle]
	\node[fill=red!50,star] (a) at (120:2cm) {B};
	\node[fill=red!50] (b) at (180:2cm) {E};
	\node[fill=red!50] (c) at (240:2cm) {E};
	\node[fill=red!50] (d) at (300:2cm) {E};
	\node[fill=red!50] (e) at (360:2cm) {E};
	\node[fill=red!50,star] (f) at (420:2cm) {B};
	
	\foreach \from/\to in {a/d, b/c, d/e, e/f, f/a}
	\draw [-] (\from) -- (\to);
	\end{tikzpicture}}
\def \corrV{\begin{tikzpicture}[scale=.4, transform shape, baseline=0pt]
	\tikzstyle{every node} = [circle]
	\node[fill=red!50,star] (a) at (120:2cm) {B};
	\node[fill=red!50] (b) at (180:2cm) {E};
	\node[fill=red!50] (c) at (240:2cm) {E};
	\node[fill=red!50] (d) at (300:2cm) {E};
	\node[fill=red!50] (e) at (360:2cm) {E};
	\node[fill=red!50,star] (f) at (420:2cm) {B};
	
	\foreach \from/\to in {a/b, b/c, c/f, d/e, f/a}
	\draw [-] (\from) -- (\to);
	\end{tikzpicture}}
\def \corrVI{\begin{tikzpicture}[scale=.4, transform shape, baseline=0pt]
	\tikzstyle{every node} = [circle]
	\node[fill=red!50,star] (a) at (120:2cm) {B};
	\node[fill=red!50] (b) at (180:2cm) {E};
	\node[fill=red!50] (c) at (240:2cm) {E};
	\node[fill=red!50] (d) at (300:2cm) {E};
	\node[fill=red!50] (e) at (360:2cm) {E};
	\node[fill=red!50,star] (f) at (420:2cm) {B};
	\path (c) edge[out=325,in=270, looseness=1.7] node {} (e);
	\foreach \from/\to in {a/b, b/d, d/f, f/a}
	\draw [-] (\from) -- (\to);
	\end{tikzpicture}}
\def \corrVII{\begin{tikzpicture}[scale=.4, transform shape, baseline=0pt]
	\tikzstyle{every node} = [circle]
	\node[fill=red!50,star] (a) at (120:2cm) {B};
	\node[fill=red!50] (b) at (180:2cm) {E};
	\node[fill=red!50] (c) at (240:2cm) {E};
	\node[fill=red!50] (d) at (300:2cm) {E};
	\node[fill=red!50] (e) at (360:2cm) {E};
	\node[fill=red!50,star] (f) at (420:2cm) {B};
	\path (b) edge[out=270,in=215, looseness=1.7] node {} (d);
	\foreach \from/\to in {a/c, c/e, e/f, f/a}
	\draw [-] (\from) -- (\to);
	\end{tikzpicture}}
\def \corrforgottenI{\begin{tikzpicture}[scale=.4, transform shape, baseline=0pt]
    \tikzstyle{every node} = [circle]
    \node[fill=red!50,star] (a) at (120:2cm) {B};
    \node[fill=red!50] (b) at (180:2cm) {E};
    \node[fill=red!50] (c) at (240:2cm) {E};
    \node[fill=red!50] (d) at (300:2cm) {E};
    \node[fill=red!50] (e) at (360:2cm) {E};
    \node[fill=red!50,star] (f) at (420:2cm) {B};
    \path (c) edge[out=325,in=270, looseness=1.7] node {} (e);
    \foreach \from/\to in {b/d, f/a}
    \draw [-] (\from) -- (\to);
    \end{tikzpicture}}

\def \corrtempxobsconn{\begin{tikzpicture}[scale=.4, transform shape, baseline=0pt]
	\tikzstyle{every node} = [circle]
	
	\node[fill=red!50,star] (a) at (90:2cm) {B};
	\node[fill=red!50] (b) at (180:2cm) {E};
	\node[fill=red!50] (c) at (270:2cm) {E};
	\node[fill=green!50,star] (d) at (360:2cm) {B};
	
	\foreach \from/\to in {a/b, b/c, c/d, d/a}
	\draw [-] (\from) -- (\to);
	\end{tikzpicture}}
\def \corrtempxobsdisc{\begin{tikzpicture}[scale=.4, transform shape, baseline=0pt]
	\tikzstyle{every node} = [circle]
	
	\node[fill=red!50, star] (a) at (90:2cm) {B};
	\node[fill=red!50] (b) at (180:2cm) {E};
	\node[fill=red!50] (c) at (270:2cm) {E};
	\node[fill=green!50, star] (d) at (360:2cm) {B};
	
	\foreach \from/\to in {a/d, b/c}
	\draw [-] (\from) -- (\to);
	\end{tikzpicture}}
\title{Impact of internal-delensing biases on searches for primordial
  $B$-modes of CMB polarisation}

\author[a]{Antón Baleato Lizancos}
\author[a,b]{Anthony Challinor}
\author[c,d]{and Julien Carron}

\affiliation[a]{Institute of Astronomy and Kavli Institute for Cosmology Cambridge, Madingley Road,\\
Cambridge, CB3 0HA, UK}
\affiliation[b]{DAMTP, Centre for Mathematical Sciences, Wilberforce Road, Cambridge, CB3 0WA, UK}
\affiliation[c]{Department of Physics \& Astronomy, University of Sussex, Brighton BN1 9QH, UK}
\affiliation[d]{Universit\'e de Gen\`eve, D\'epartement de Physique Th\'eorique et CAP, 24 Quai Ansermet, CH-1211 Gen\`eve 4, Switzerland}
\emailAdd{a.baleatolizancos@ast.cam.ac.uk}
\emailAdd{a.d.challinor@ast.cam.ac.uk}
\emailAdd{julien.carron@unige.ch}

\abstract{Searches for the imprint of primordial gravitational waves
  in degree-scale CMB $B$-mode polarisation data must account for
  significant contamination from gravitational lensing. Fortunately,
  the lensing effects can be partially removed by combining
  high-resolution $E$-mode measurements with an estimate of the
  projected matter distribution. In the near future, experimental
  characteristics will be such that the latter can be reconstructed
  internally with high fidelity from the observed CMB, with the
  $EB$ quadratic estimator providing a large fraction of the
  signal-to-noise. It is a well-known phenomenon in this context that
  any overlap in modes between the $B$-field to be delensed and the
  $B$-field from which the reconstruction is derived leads to a
  suppression of delensed power going beyond that which can be
  attributed to a mitigation of the lensing effects. More importantly,
  the variance associated with this spectrum is also
  reduced, posing the question of whether the additional power
  suppression could help better constrain the tensor-to-scalar ratio,
  $r$. In this paper, we show this is not the case, as suggested but not quantified in previous work.
  We develop an analytic model for the biased delensed $B$-mode angular power spectrum, which suggests a simple renormalisation prescription to avoid bias on the inferred tensor-to-scalar ratio. With this approach, we learn that the bias necessarily leads to a \emph{degradation} of the signal-to-noise on a primordial component compared to ``unbiased delensing''. Next, we assess the impact of removing from the lensing reconstruction any overlapping $B$-modes on our ability to constrain $r$, showing that it is in general advantageous to do this rather than modelling or renormalising the bias. Finally, we verify these results within a maximum-likelihood inference framework applied to simulations.}

\begin{document}
\maketitle
\flushbottom

\section{\label{sec:introduction}Introduction}
    The theoretical framework best fitting cosmological observations includes a period of accelerated expansion at very early times called \textit{inflation}. It is a general feature amongst inflationary models that a background of gravitational waves (tensor perturbations) would have been generated during that era along with fluctuations in the density (scalar perturbations). Although difficult to be detected directly, primordial gravitational waves should have left an imprint on the temperature and polarisation of the cosmic microwave background (CMB) that may be detectable if inflation occurred at a sufficiently high energy scale~\cite{Polnarev:1985,Kamionkowski:1996zd,Seljak:1996gy}.

    Although scalar perturbations are now known to overwhelm any contribution from tensor perturbations to the temperature anisotropy ($T$)  or the gradient-like $E$-mode component of the polarisation, it is possible to form a curl-like component ($B$-mode) that at recombination is sourced only by tensor perturbations (in linear theory). A detection of a primordial $B$-mode is therefore widely regarded as a conclusive test of inflation. Furthermore, a measurement of the ratio of primordial tensor-to-scalar power, $r$, would reveal the energy scale of inflation. Current experimental bounds place $ r<0.06 $ with $95\,\%$ confidence~\cite{ref:bicep2_18}.

    As we search for a signal below this level, it is not sufficient just to improve the sensitivity of experiments, for gravitational lensing of CMB photons by the large-scale distribution of matter in the Universe contaminates the signal by converting part of the primordial $E$-mode signal into $B$-mode~\cite{Zaldarriaga:1998ar}. The $B$-modes induced by gravitational lensing were first detected by~\cite{ref:hanson_13}. These have a power spectrum resembling that of white noise with $\Delta_{P}=5 \mu \text{K\,arcmin}$ on the large angular scales where the primordial signal is expected to be strongest. Consequently, the uncertainty on measurements of the primordial tensor signal will henceforth be dominated by the large variance associated with the lensing component unless the latter can be removed.

    Unfortunately, the removal of lensing $B$-modes cannot be achieved via multi-frequency cleaning, since lensing merely induces a remapping of photons by small deflection angles (of order a few arcminutes) while preserving the blackbody spectrum of the unlensed CMB. Progress can be made, however, by estimating the projected matter distribution responsible for the deflections, and undoing them (at least partially). Several groups have already applied this procedure, known as \textit{delensing}, to real $B$-mode data~\cite{ref:carron_17, ref:spt_17, Planck2018:lensing, ref:polarbear_delensing_19, ref:han_20, ref:bicep_delensing}. For the foreseeable future (as long as lensing residuals remain larger than $O(1\%)$ of the original amplitude), remapping by the inverse deflections can be approximated by the more analytically transparent and computationally efficient subtraction of a template estimating the lensing $B$-modes to leading order~\cite{ref:template_B_paper_in_prep}, that is, a weighted convolution of $E$-mode observations with some estimate of the lensing deflections on the sky.  This procedure, which we refer to as \emph{template delensing}, has successfully been applied to data~\cite{ref:planck_template, ref:spt_17}.

    The extent to which the lensing signal can be removed depends on the fidelity with which the lensing potential can be estimated. For sufficiently low experimental noise levels, the lensing potential, $\phi$, can be reconstructed internally from the CMB maps themselves by employing either the quadratic estimators of ref.~\cite{ref:okamoto_hu_2003} or the more powerful, albeit analytically and computationally complex, maximum-likelihood~\cite{ref:hirata_03_polarization} or Bayesian methods~\cite{ref:carron_17_maximum, ref:millea_17}. For experimental noise levels available already in the next generation of CMB observatories, a quadratic combination involving $E$- and $B$-fields will provide a sizeable fraction of the signal-to-noise, dominating over other estimators in the regime where $\Delta_{P}<5\,\mu\text{K\,arcmin}$.

     However, ref.~\cite{ref:teng_11} showed that whenever there is an overlap in modes between the field we wish to delens and the fields from which a lensing reconstruction is derived, the delensed power is subject to a bias that leads to a suppression in power going beyond that which can be attributed to a mitigation of the lensing effects. More importantly, the variance associated with this delensed spectrum is also reduced, posing the question of whether the bias could help better constrain the tensor-to-scalar ratio, $r$.

     In this paper, we set out to understand the angular power spectrum of $B$-modes after internal delensing with an $EB$ quadratic estimator, as a function of experimental characteristics and, importantly, primordial $B$-mode power, assuming no foreground or survey non-idealities. In section~\ref{sec:lensing_reconstruction}, we introduce the quadratic estimators for lensing reconstruction, focusing on the $EB$ combination. In section~\ref{sec:template_delensing}, we show how an estimate of the lensing $B$-modes at the map level can be made by combining observations of $E$ and estimates of $\phi$. Then, in section~\ref{sec:delensed_ps}, we introduce analytic models for the angular power spectrum of delensed $B$-modes in both the cases with and without overlapping modes, and compare them to simulations. (These models are derived carefully in appendix~\ref{appendixa}, and the simulations are described in appendix~\ref{appendix:sims}.)
     In section~\ref{sec:covariances}, we consider simple models for the covariance of the delensed $B$-mode power spectrum. We combine the power spectrum models and covariances in a maximum-likelihood framework in section~\ref{sec:likelihood}, where we simulate inferences of $r$ to compare quantitatively the different ways in which the bias can be dealt with. Finally, in section~\ref{sec:obs_x_del_delensing}, we briefly study the impact of the bias on an alternative estimator which correlates observed and delensed $B$-modes.  

    The results in this paper will often refer to the experimental specifications of the upcoming Simons Observatory (SO)\cite{ref:so_science_paper}, which will feature a large-aperture telescope (LAT) responsible, among other scientific targets, for lensing reconstruction; and separate small-aperture telescopes (SATs) for observations of $B$-modes on large angular scales. However, the insights developed here apply more widely, and will likely be relevant to any application of internal delensing that uses information from the $B$-modes for the purpose of lensing reconstruction.

\section{\label{sec:lensing_reconstruction}Quadratic estimators for lensing reconstruction}
    In order to undo the deflections induced by lensing, an estimate of the projected matter distribution on the sky -- which determines the lensing potential -- is required. For the sensitivities and resolution of current CMB experiments, the best possible estimate is obtained from tracers external to the CMB such as the cosmic infrared background (CIB) or very deep galaxy surveys \cite{ref:smith_12_external,ref:sherwin_15,ref:larsen_16,Schmittfull:2017ffw}. The CIB retains a high degree of correlation with the smaller-scale lenses at high redshift that are important for converting $E$-modes into $B$-modes,
    something that is not yet possible with internal lensing reconstruction, which is very noisy on the relevant angular scales.
    However, this situation will change with planned CMB polarisation surveys, which can provide high signal-to-noise lensing reconstructions on nearly all scales relevant for $B$-mode delensing~\cite{ref:s4}. Consequently, reconstructions of $\phi$ derived from the CMB fields themselves will ultimately provide the best delensing performance.

    Internal reconstruction techniques use the fact that, if we could average over realisations of the CMB while keeping the lensing potential fixed, lensing would break statistical isotropy by inducing correlations between fluctuations on different scales; consequently, the lensing potential can be reconstructed by combining many off-diagonal correlations. In general, the optimal internal reconstruction of $\phi$ will be obtained from studying the likelihood function of the lensed CMB temperature and polarisation anisotropies~\cite{ref:hirata_03_temperature, ref:hirata_03_polarization, ref:carron_17_maximum, ref:millea_17}. However, for the noise levels attainable in the near future, the optimal lensing reconstruction has been shown (see, e.g., ref.~\cite{ref:core_lensing}) to be equivalent to that from the more tractable and computationally efficient ``quadratic estimators'' of refs.~\cite{ref:hu_okamoto_2002, ref:okamoto_hu_2003}. In fact, in the upcoming era of high-resolution CMB experiments such as the Simons Observatory \citep{ref:so_science_paper} and SPT3G \cite{ref:spt3g_14}, which feature experimental noise levels $1\,\mu\mathrm{K\, arcmin}<\Delta_{P}<10\,\mu\mathrm{K\, arcmin}$, the optimal reconstruction will arise from a combination of external tracers and internal reconstructions involving quadratic estimators~\cite{ref:sherwin_15}.

\begin{figure}
        \centering
        \includegraphics[width=0.7\textwidth]{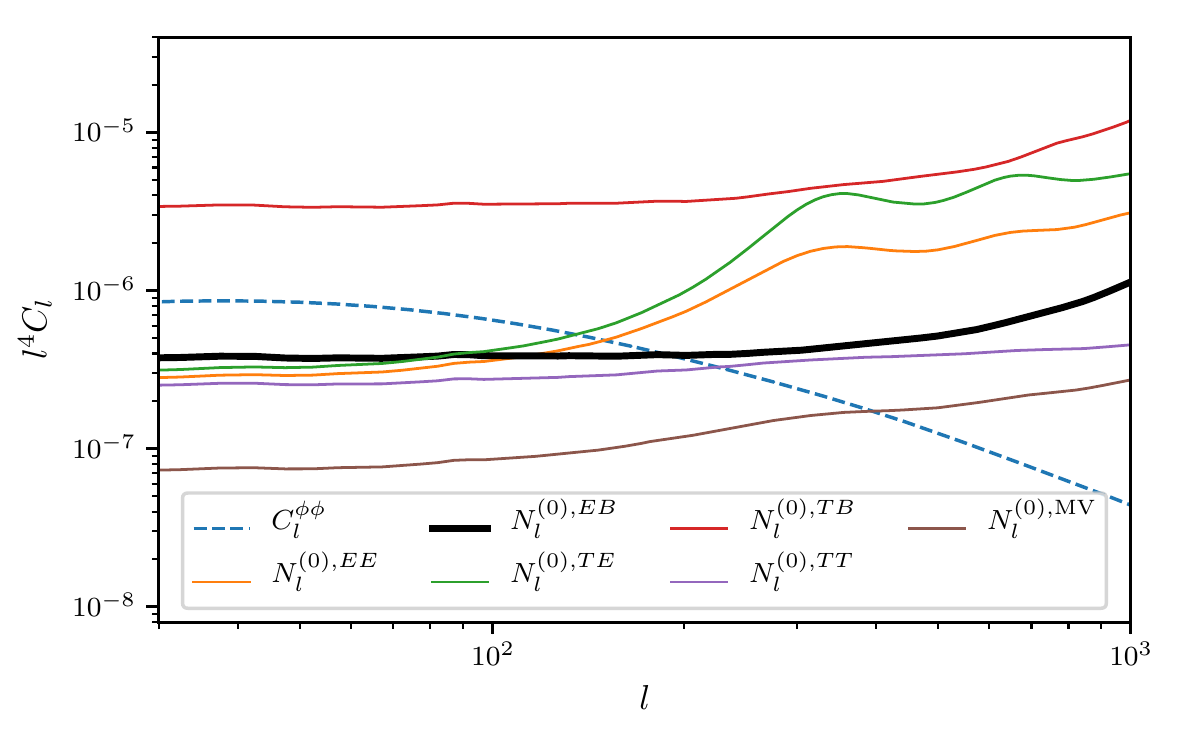}
        \caption{Lensing reconstruction noise levels associated with the five different quadratic estimators, and their minimum-variance combination, for an experiment with resolution $\theta_{\mathrm{FWHM}}=1.5\,\text{arcmin}$ (full width at half maximum) and isotropic white noise with $\Delta_{T}=6\, \mu \text{K\,arcmin}$ for temperature and $\Delta_{P}= \sqrt{2}\Delta_{T}$ for polarisation. The maximum multipole used in the reconstruction is $l_{\mathrm{max}}=3000$. These specifications are similar to the goals for the SO LAT at 145\,GHz. The dashed blue line is the lensing potential power spectrum.}
        \label{fig:reconstruction_noise_levels}
    \end{figure}

    A minimum-variance internal reconstruction of the lensing potential can be obtained from combining different quadratic estimators, as seen in figure~\ref{fig:reconstruction_noise_levels}. For low enough noise levels, the quadratic estimator involving $E$ and $B$ fields is expected to provide the highest signal-to-noise reconstruction -- the reason being that, in this case, the dominant source of reconstruction noise involves the Gaussian contraction $\langle EE\rangle \langle BB\rangle$, and the lensing and primordial contributions to $\langle BB\rangle$ are small~\cite{ref:core_lensing}. It is because of its relevance in upcoming efforts to delens the CMB that we focus on the $EB$ estimator in this paper. It takes the form
    \begin{equation}\label{eqn:eb_est}
    \hat{\phi}_{EB}(\bm{L}) = A^{EB}_{L} \int \meas W(\bm{L} - \bl,  - \bl)   \frac{\tilde{C}^{EE, \, \mathrm{fid}}_{l}}{C^{EE,\,\obs,\,\fid,\,\lat }_{l}C^{BB,\,\obs,\,\fid,\,\lat }_{|\bl-\bm{L}|}}   E^{\obs,\lat}(\bm{l})B^{\obs,\lat}(\bm{L}-\bm{l}),
    \end{equation}
    where $E^{\obs,\lat}$ and $B^{\obs,\lat}$ are beam-deconvolved observed fields\footnote{In the case where an observatory consists of separate telescopes with different characteristics and mapping strategies, the one with the larger aperture -- labeled here as $\text{LAT}$, in contrast with the $\text{SAT}$, or small aperture telescope -- will provide the superior lensing reconstruction as it can more finely resolve the small scales where lensing dominates.}, $A^{EB}_{L}$ is a normalisation factor, $C^{XX,\,\obs,\,\fid,\,\lat }_l$ is the fiducial, total power observed by the LAT which is used to inverse-variance filter\footnote{For simplicity, we take this filter to be diagonal. In practical applications with masking and/or inhomogeneous noise, the optimal filtering function will no longer be diagonal.} the field $X$, and the geometric coupling
    \begin{equation}\label{eqn:w_def}
    W(\bm{l}, \bm{l}') \equiv \bm{l}' \cdot (\bm{l}-\bm{l}') \sin 2\left( \psi_{\bm{l}} - \psi_{\bm{l}'} \right) .
    \end{equation}
    Here, $\psi_{\bm{l}}$ is the angle between $\bm{l}$ and the $x$-axis used to define positive Stokes parameter $Q$, and similarly for $\psi_{\bm{l}'}$.
    Notice the use of the fiducial lensed $\tilde{C}_l^{EE,\fid}$ instead of its unlensed counterpart, since this has been shown to optimise the correlation of estimates of the lensing potential with the underlying truth\footnote{For the experimental configurations considered in this paper, the delensing performance obtained by using these weights appears to be indistinguishable from the case where the unlesed spectra are used instead.}~\cite{ref:hanson_11,ref:lewis_11}. If eq.~\eqref{eqn:eb_est} were to minimise exactly the variance of $\hat{\phi}_{EB}$, it would include an additional term proportional to the primordial $B$-mode spectrum\footnote{While we denote as $\tilde{C}_l^{EE}$ the lensed $E$-mode spectra (including primordial $E$-modes), we follow the standard convention that $\tilde{C}_l^{BB}$ comprises only the lensing contribution to the $B$-mode spectrum, with the primordial part given separately by $C_l^{BB,t}$}, $C_l^{BB,t}$, like the one shown but replacing
    $W(\bm{L}-\bm{l},-\bm{l})\tilde{C}_l^{EE,\mathrm{fid}} \rightarrow -W(-\bm{l},\bm{L}-\bm{l})C_{|\bm{L}-\bm{l}|}^{BB,\mathrm{t,fid}}$.
    We ignore it here, however, as its effect is negligible for values of $r$ compatible with experimental bounds. 

\section{\label{sec:template_delensing}Template delensing of the $B$-mode }
    Once we have an estimate for the lensing potential, delensing can be performed by using it to build a template approximating the lensing $B$-modes,  $B^{\temp}$, which can then be subtracted from the observed $B$-modes,\footnote{In multi-telescope observatories, small aperture telescopes (SATs), which have proven stability to large-scale signals, are being adopted to pursue primordial $B$-modes.} $B^{\obs, \,\sat}$, to yield a delensed field
    \begin{equation}\label{eqn:b_del}
        B^{\text{del}} = B^{\obs,\,\sat} -B^{\temp}.
    \end{equation}
    To leading order in $\boldsymbol{\nabla} \phi$ (what is usually referred to as the ``gradient approximation'', and is known to be an excellent approximation to the non-perturbative calculation of the lensed $B$-mode spectrum on large angular scales \cite{ref:challinor_2005}), the template can be constructed as
    \begin{equation}\label{eqn:template}
    B^{\text{temp}}(\bm{l}) = \int \measp f(\bm{l}, \bm{l}') W(\bm{l}, \bm{l}') E^{\obs,\lat}(\bm{l}')\hat{\phi}(\bm{l}-\bm{l}'),
    \end{equation}
    with $W(\bm{l}, \bm{l}')$ as defined in eq.~\eqref{eqn:w_def}. The function $f(\bm{l}, \bm{l}')$ can be chosen to minimise the power spectrum of the delensed field~\eqref{eqn:b_del}, which ref.~\cite{ref:smith_12_external} shows to be the case when the fields are filtered with
    \begin{equation}
        f(\bm{l}, \bm{l}') =  \We_{l'} \Wp_{|\bl - \bl'|},
    \end{equation}
     where
    \begin{equation}\label{eqn:wiener_filters}
    \We_l\equiv \frac{\tilde{C}_l^{EE, \fid}}{C_l^{EE,\text{\obs,}\fid,\,\lat}} \quad \text{and} \quad
    \Wp_L \equiv \frac{{C}_L^{\phi\phi, \fid}}{{C}_L^{\phi\phi, \fid}+N_L^{\phi\phi}} ,
    \end{equation}
    are Wiener filters for the $E$-modes and the estimate of the lensing potential. Here, $ N_L^{\phi\phi} $ is the reconstruction noise on $ \hat{\phi} $. In practice, we use $N_L^{\phi\phi} \approx N^{(0),EB}_L$, the disconnected (Gaussian) contribution to the power spectrum of $\hat{\phi}$,
    and ignore higher-order terms such as $N^{(1),EB}_L$, as the former dominates over the latter by several orders of magnitude on all relevant scales~\cite{ref:cooray_kesden_03}. Notice also that eq.~\eqref{eqn:template} uses the observed (lensed and noisy) $ E^{\obs, \lat}$, rather than unlensed (or delensed) $E$-modes. This is to be preferred \emph{for template delensing} since an approximate cancellation suppresses higher-order contributions in $C_L^{\phi\phi}$ to the delensed power spectrum, which would otherwise be as big as 30\,\% on large scales~\citep{ref:template_B_paper_in_prep}.

\section{\label{sec:delensed_ps}Angular power spectrum of delensed $B$-modes}
    Most inflationary models predict a spectrum of primordial perturbations that are very nearly Gaussian distributed. This character extends approximately to the fluctuations of the primary CMB, as linear-theory transfer functions are an excellent approximation in the early universe. For such Gaussian fields, the data can be losslessly compressed into a power spectrum, making the latter one of the most powerful tools for studying the CMB. Although the \textit{lensed} CMB is actually non-Gaussian (as a result of the Gaussian primary CMB fields being displaced by the nearly-Gaussian lensing potential), delensing has been shown to mitigate its non-Gaussian character to the point where the delensed $B$-modes -- if delensed in such a way that the $EB$ delensing bias is avoided -- deviate only slightly from Gaussianity~\cite{ref:namikawa_15} and the information lost from working with the power spectrum is relatively small.

    More importantly, in practical applications involving masked observations and anisotropic noise, any computation of the exact likelihood of the data becomes intractable and one must work instead with approximate forms, most typically involving estimators of the power spectra and their covariance~\cite{ref:hamimeche_2008}.

    For these reasons, most efforts to detect primordial $B$-modes work with the power spectrum of the delensed field of eq.~\eqref{eqn:b_del}:
    \begin{align}\label{eqn:contributions_to_delensed_spectrum}
    \langle B^{\text{del}}(\bl_1)B^{\text{del}}(\bl_2)\rangle & =  \langle B^{\obs,\,\sat}(\bl_1) B^{\obs,\,\sat}(\bl_2)\rangle 
    -2\langle B^{\temp}(\bl_1) B^{\obs,\,\sat}(\bl_2)\rangle \nonumber\\
    &\qquad + \langle B^{\temp}(\bl_1) B^{\temp}(\bl_2)\rangle \nonumber\\
    & \equiv (2\pi)^2\delta^{(2)}(\bl_1+\bl_2) C_{l_1}^{BB,\text{del}},
    \end{align}
    where in the last line we have used the statistical isotropy of the CMB to define $C_l^{BB,\text{del}}$, the angular power spectrum of the delensed $B$-modes. In the remainder of this section we examine this expression in detail. In section~\ref{sec:unbiased_del_ps}, we evaluate it in the case where the statistical errors in the estimated lensing potential
    are independent of the lensed CMB fields, exploring the reduction in power associated with removal of part of the lensing signal. Then, in section~\ref{sec:biases}, we study eq.~\eqref{eqn:contributions_to_delensed_spectrum} in the case where $\hat{\phi}$ is derived from an $EB$ quadratic estimator; new couplings then appear that further suppress the delensed power spectrum beyond a simple removal of lensing power. We conclude the section by proposing an analytic model to capture the behaviour of such ``bias'' terms.

	\subsection{The unbiased case: reconstruction errors statistically independent of the lensed CMB\label{sec:unbiased_del_ps}}
    If the noise on $\hat{\phi}$ were independent of the $B$-mode we would like to delens, as would be the case if an external tracer were used for $\hat{\phi}$, the delensed power spectrum would take the form
    \begin{equation}\label{eqn:unbiased_del_ps}
    C_l^{BB,\text{del}, \text{unbiased}}= N_l^{BB,\sat} + C_l^{BB,t} + C_l^{BB,\text{res}}.
    \end{equation}
    In addition to the instrumental noise, $ N_l^{BB,\,\sat} $, and primordial component, $C_l^{BB,\mathrm{t}}  $, there is a contribution from residual deflections -- imperfect delensing -- given to leading order in lensing by \cite{ref:smith_12_external}
    \begin{align}\label{eqn:naive_clbbres}
    C_l^{BB,\text{res}} &\approx \int \measp W^2(\bm{l}, \bm{l}')C_{l'}^{EE}C_{|\bm{l} - \bm{l}'|}^{\phi\phi} \big[1-\We_{l'} \Wp_{|\bm{l}-\bm{l}'|}\big]\nonumber\\
    &=\tilde{C}_l^{BB} - C_l^{W},
    \end{align}
    where we have defined
    \begin{equation}\label{eqn:def_clbbW}
    C_{l}^{W} \equiv \int \measp W^2(\bl, \bl')\big[\We_{l'} C_{l'}^{EE, \fid}\big]\big[\Wp_{|\bl-\bl'|}C^{\phi\phi, \mathrm{fid}}_{|\bl-\bl'|}\big]
    \end{equation}
    and assumed that our fiducial model for the lensing power spectrum, $C^{\phi\phi, \mathrm{fid}}_l$, is correct. We note that $C_l^W$ is simply the fiducial power spectrum of the $B$-mode template.
    In appendix~\ref{appendix:std_calc_from_full}, we explain how eq.~\eqref{eqn:naive_clbbres} can be recovered in the analytic framework developed in section~\ref{sec:biases} to characterise the delensing bias. Notice that, in the limit of no observational noise and a perfect $\hat{\phi}$, we have $\We_l \rightarrow 1$ and $\Wp_l \rightarrow 1$ and all of the (leading-order) lensing signal is removed. Higher-order terms mean that template delensing can, in fact, be used to reduce the lensing contribution to the power spectrum to approximately $1\,\%$ of its original level. This is in the case where lensed $E$-modes are employed in the construction of the $B$-mode template -- the alternative of using unlensed/delensed $E$-modes performs worse with a lensing residual of order $30\,\%$ as noted above~\citep{ref:template_B_paper_in_prep}. Were estimates of the lensing potential accurate enough to reduce the delensed power to the $1\,\%$ level, template delensing should be replaced with non-perturbative methods whereby the Wiener-filtered $\hat{\phi}$ is used to remap the observed CMB directly (see, e.g., \cite{ref:larsen_16, ref:carron_17, ref:polarbear_delensing_19, Planck2018:lensing} for demonstrations of this method).

    Henceforth, we will refer to eq.~\eqref{eqn:unbiased_del_ps} as the \emph{unbiased} delensed power spectrum in order to differentiate it from the case where the errors in $\hat{\phi}$ are statistically dependent on the lensed CMB, as is the case when internal delensing with overlapping modes. We now consider the new terms that arise in the delensed power for this latter case.
  
    \subsection{The biased case: $\hat{\phi}$ obtained from an $EB$ quadratic estimator}\label{sec:biases}
    We discussed in section~\ref{sec:lensing_reconstruction} that, in the upcoming era of low-noise CMB experiments, the highest signal-to-noise reconstruction using quadratic estimators will arise from combining $E$ and $B$ fields. Crucially, ref.~\cite{ref:teng_11} first noticed that, if such a reconstruction is used to delens observations of $ B $-modes, the delensed power spectrum will be biased: while eq.~\eqref{eqn:naive_clbbres} continues to quantify the power spectrum due to residual deflections, the total delensed spectrum will show an additional suppression of power beyond that which can be attributed to delensing. In this section, we add to the work of \cite{ref:teng_11,ref:namikawa_nagata_14,ref:namikawa_17} by providing a detailed calculation of the biased delensed $B$-mode power spectrum, including terms neglected by eq.~\eqref{eqn:naive_clbbres}.

    In order to isolate the effects of the bias, we assume henceforth that the lensing reconstruction is obtained exclusively from an $EB$ quadratic estimator. Although this is set to be the dominant source of information on lensing for upcoming experiments, in real applications the optimal reconstruction will, in fact, arise as a minimum-variance (MV) combination of several quadratic estimators. In that scenario, the bias arising from $EB$ reconstruction needs to be propagated through the co-adding procedure, and will ultimately be reduced relative to the case where only the $EB$ estimator is used because the other estimators it is combined with do not source a bias. Hence, to obtain the amplitude of the bias on the delensed $B$-mode spectrum and on $r$ in the MV case, the results quoted in this paper would need to be scaled appropriately by the MV weight pertaining to the $EB$ estimator.

    In this \emph{biased} case, the first term in the delensed $B$-mode spectrum of eq.~\eqref{eqn:contributions_to_delensed_spectrum} is still
    \begin{equation}\label{eqn:obs_x_obs}
    \langle B^{\obs,\,\sat}(\bl_1) B^{\obs,\,\sat}(\bl_2)\rangle = (2\pi)^2\delta^{(2)}(\bm{l}_1+\bm{l}_2)C_{l_1}^{BB,\text{obs,\,\sat}},
    \end{equation}
    but the other two correlators receive important new contributions. Substituting in eqs.~\eqref{eqn:eb_est} and~\eqref{eqn:template}, the second term becomes
    \begin{align}\label{eqn:temp_x_obs}
    \langle B^{\temp}(\bl_1) B^{\obs,\, \sat}(\bl_2)\rangle = &\int \frac{d^2\bl'_1 d^2\bl''_1}{(2\pi)^4} \We_{l'_1} \Wp_{|\bl_1-\bl'_1|} \nonumber\\
    & \times \frac{A^{EB}_{|\bl_1-\bl'_1|} \tilde{C}^{EE, \, \mathrm{fid}}_{l''_1}}{C^{EE,\,\obs,\,\fid,\,\lat }_{l''_1}C^{BB,\,\obs,\,\fid,\,\lat }_{|\bl_1-\bl'_1-\bl''_1|}} W(\bl_1,\bl'_1) W(\bl_1-\bl'_1-\bl''_1, -\bl''_1) \nonumber\\
    &\times \langle E^{\obs,\,\lat}(\bl'_1) E^{\obs,\,\lat}(\bl''_1) B^{\obs,\,\lat}(\bl_1-\bl'_1-\bl''_1)B^{\obs,\,\sat}(\bl_2)\rangle.
    \end{align}
    The last term, which correlates two templates, takes the form
    \begin{align}\label{eqn:temp_x_temp}
    \langle B^{\temp}(\bl_1) B^{\temp}(\bl_2)\rangle  = &\int\frac{d^2\bl'_1 d^2\bl'_2 d^2\bl''_1 d^2\bl''_2}{(2\pi)^8}\We_{l'_1}\We_{l'_2}\Wp_{|\bl_1-\bl'_1|}\Wp_{|\bl_2-\bl'_2|}\nonumber\\
    & \times \frac{A^{EB}_{|\bl_1-\bl'_1|}A^{EB}_{|\bl_2-\bl'_2|}}{C^{BB,\,\obs,\,\fid,\,\lat }_{|\bl_1-\bl'_1-\bl''_1|}C^{BB,\,\obs,\,\fid,\,\lat }_{|\bl_2-\bl'_2-\bl''_2|}}\frac{\tilde{C}^{EE, \, \mathrm{fid}}_{l''_1}}{C^{EE,\,\obs,\,\fid,\,\lat }_{l''_1}}\frac{\tilde{C}^{EE, \, \mathrm{fid}}_{l''_2}}{C^{EE,\,\obs,\,\fid,\,\lat }_{l''_2}}\nonumber\\
    & \times W(\bl_1,\bl'_1)W(\bl_2,\bl'_2)W(\bl_1-\bl'_1-\bl''_1, -\bl''_1) W(\bl_2-\bl'_2-\bl''_2, -\bl''_2)\nonumber\\
    &\times \langle E^{\obs,\,\lat}(\bl'_1) E^{\obs,\,\lat}(\bl''_1) B^{\obs,\,\lat}(\bl_1-\bl'_1-\bl''_1) \nonumber\\
    &\qquad \times E^{\obs,\,\lat}(\bl'_2) E^{\obs,\,\lat}(\bl''_2)B^{\obs,\,\lat}(\bl_2-\bl'_2-\bl''_2)\rangle.
    \end{align}

    The evaluation of the four- and six-point functions that appear in eqs~\eqref{eqn:obs_x_obs} and~\eqref{eqn:temp_x_temp} is discussed in detail in appendix~\ref{appendixa}. These are expanded in terms of connected $n$-point functions with $n=2$ and $4$ (the connected six-point function is higher order in $C_l^{\phi\phi}$). A subset of these terms combine to give the standard unbiased result~\eqref{eqn:naive_clbbres}, as shown in appendix~\ref{appendix:std_calc_from_full}. The remaining terms introduce corrections, with the most important of these identified and evaluated in appendix~\ref{appendix:calculation_of_eb_bias}. Combining these results, we show that the biased delensed $B$-mode power spectrum can be approximated as
    \begin{equation}\label{eqn:biased_delensed_spectrum_model}
        C_l^{BB,\text{del}} = (C_l^{BB,\text{res}} + C_l^{BB,t})(D_l-1)^2 + D_l^2C_l^W+ N_l^{BB,\sat} +  N_l^{BB,\lat}D_l^2 - 2D_l N_l^X,
    \end{equation}
    where the correlation of experimental noise in the SAT and the LAT is denoted as
    \begin{equation}
        N_l^X = \begin{cases}
        0 & \text{if SAT and LAT are separate,}\\
        N_l^{BB,\lat}=N_l^{BB,\sat} & \text{for a single telescope.}
        \end{cases}
    \end{equation}
    We have also defined
    \begin{equation}\label{eqn:D_l}
    D_{l} \equiv \frac{1}{C_{l}^{BB,\obs,\fid,\lat}} \int \measp W^2(\bl,\bl')\big[\We_{l'}\tilde{C}_{l'}^{EE}\big]\big[\Wp_{|\bl-\bl'|}A_{|\bl-\bl'|}^{EB}\big] ,
    \end{equation}
    whose origin and properties are discussed further below.
    
    Equation~\eqref{eqn:biased_delensed_spectrum_model} captures the general case where a large-aperture telescope (LAT) focuses on lensing reconstruction while a separate small-aperture telescope (SAT) makes observations of the $B$-modes on large angular scales. In such a configuration, the experimental noise is uncorrelated between instruments and $N_l^{X}=0$. On the other hand, the case where a single telescope is used for both purposes can easily be recovered by letting $N_l^{X}=N_l^{BB,\lat}=N_l^{BB,\sat}$, in which case the biased delensed spectrum reduces to
    \begin{equation}\label{eqn:biased_delensed_spectrum_model_single_tel}
       C_l^{BB,\text{del}} = (C_l^{BB,\text{res}} + C_l^{BB,t} + N_l^{BB})(D_l-1)^2 + D_l^2C_l^W.
    \end{equation}
    
    All of the correction terms in eq.~\eqref{eqn:biased_delensed_spectrum_model} are proportional to one or two powers of $D_l$; for $D_l=0$ it reduces to the unbiased result~\eqref{eqn:naive_clbbres}. The function $D_l$ arises when one contracts the $B$-mode template over the pair of observed $E$-modes that enter explicitly. In detail,
    \begin{equation}
    \langle B^{\temp}(\bl) \rangle_{E^{\obs,\lat}} = B^{\obs,\lat}(\bl) D_l \, .  
    \label{eqn:originoflocalbias}
    \end{equation}
    All of the correction terms retained in eq.~\eqref{eqn:biased_delensed_spectrum_model} arise from such contractions. For example, since $E$- and $B$-modes are uncorrelated, the disconnected contribution to $\langle B^{\temp}(\bl_1) B^{\obs,\,\sat}(\bl_2)\rangle$ gives a term
    \begin{align}
    \langle B^{\temp}(\bl_1) B^{\obs,\,\sat}(\bl_2)\rangle &\supset D_{l_1} \langle B^{\obs,\lat}(\bl_1) B^{\obs,\,\sat}(\bl_2)\rangle   \nonumber \\
    &= (2\pi)^2 \delta^{(2)}(\bm{l}_1+\bm{l}_2) D_{l_1}\left(\tilde{C}^{BB}_{l_1}+{C}^{BB,\mathrm{t}}_{l_1}+{N}^{X}_{l_1}\right) \, .
    \label{eqn:dominantbias}
    \end{align}
    Note how this involves the cross-correlation of the experimental noise between the SAT and the LAT, $N_l^X$, and the tensor $B$-mode power. It is the main correction term and, given that it enters the delensed power through $-2 \langle B^{\temp}(\bl_1) B^{\obs,\,\sat}(\bl_2)\rangle$, \emph{suppresses} the power further. (The other correction terms we retain increase the delensed power.)
    
    \begin{figure}
        \centering
        \includegraphics[width=0.7\textwidth]{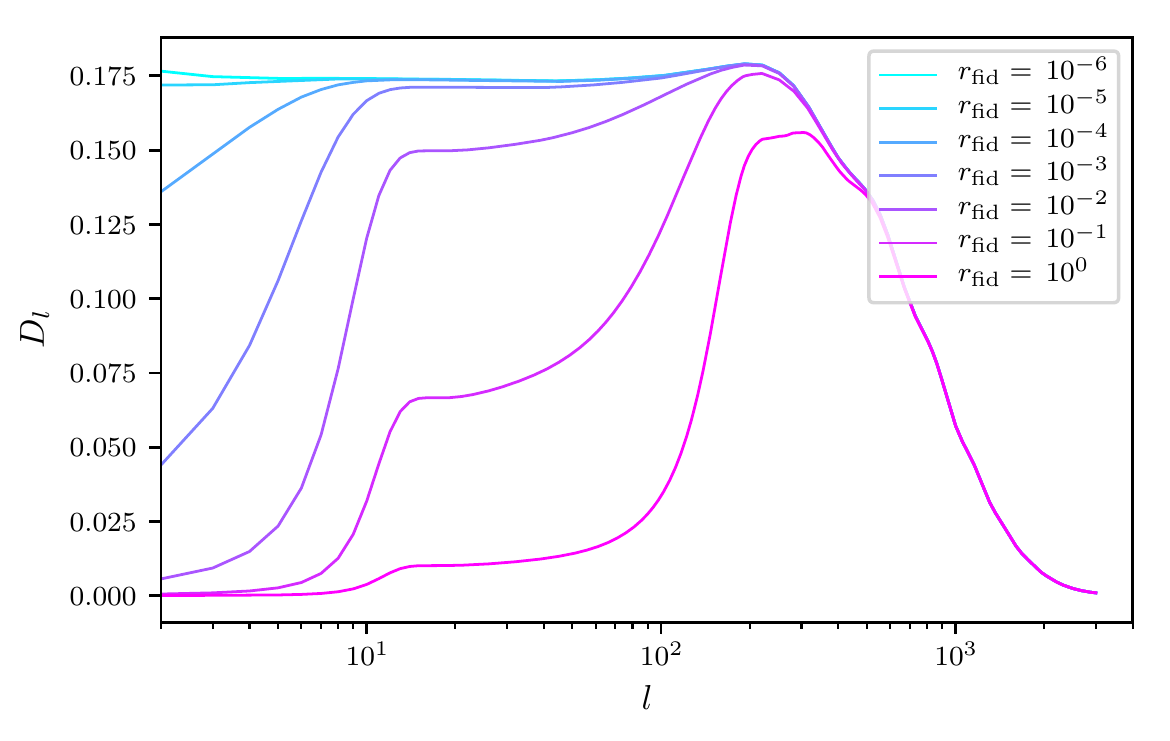}
        \caption{Evaluation of $D_l$, defined in eq.~\eqref{eqn:D_l}, for several values of the fiducial $r_{\mathrm{fid}}$ in the filter $1/C_{l}^{BB,\obs,\fid,\lat}$ applied to the $B$-modes in lensing reconstruction with the $EB$ quadratic estimator. The experimental set-up is as in figure~\ref{fig:reconstruction_noise_levels}.}
        \label{fig:Dl_values}
    \end{figure}

    We can gain further insight into $D_l$ by noting that in the limit of noiseless $E$-mode observations, and setting the normalisation of the quadratic estimator to $A_L^{EB}=N^{(0),EB}_L$ (which is correct in the usual limit that the fiducial spectra used to inverse-variance filter the CMB fields in the lensing reconstruction are close to the true total power), $D_l \rightarrow C_l^{BB, \mathrm{res}}/C_{l}^{BB,\obs,\fid,\lat}$ [cf.\ eq.~\eqref{eqn:naive_clbbres}]. Consequently, in this limit $D_l$ can be interpreted as the ratio of residual lensing power to fiducial power used to inverse-variance filter the $B$-modes in the lensing reconstruction. More generally, the Wiener filter $\We_l$ will reduce $D_l$ below this value. In all cases $0 < D_l < 1$. In figure~\ref{fig:Dl_values}, we see that $D_l$ depends strongly on large scales on the fiducial primordial $B$-mode power contained in the inverse-variance filters and parametrised by $r_{\mathrm{fid}}$. (The figure includes large values of $r_{\mathrm{fid}}$ inconsistent with observations but included for illustration.)
    To understand this behaviour recall that, for $r\sim 0.01$, $C_l^{BB,\mathrm{t}}$ becomes comparable to the lensing power on large angular scales between $10<l<100$. By raising $r_{\mathrm{fid}}$ above this value, we are effectively down-weighting the large-scale lensing $B$-modes in the fields from which the reconstruction is derived.  Although this will not significantly affect the fidelity of the lensing reconstruction -- which is derived from information coming chiefly from smaller angular scales -- it will give a sizeable reduction in $D_l$ and the bias in the total delensed power on large scales.

    The result in eq.~\eqref{eqn:biased_delensed_spectrum_model} is a generalisation of that in ref.~\cite{ref:namikawa_nagata_14} (eq.~A16 there), which assumes $r=0$ and separate LAT and SAT observations ($N_l^X=0$). An important insight from our result is that the bias term responsible for the additional suppression of power, eq.~\eqref{eqn:dominantbias}, is proportional to the \emph{total} observed $B$-mode power, and not simply its lensing component\footnote{The model of eq.~\eqref{eqn:biased_delensed_spectrum_model_single_tel} (with $N_l^{X}=0$) can be obtained from eq.~(A.16) of ref.~\cite{ref:namikawa_nagata_14} by adding $-2D_lC_l^{BB, \mathrm{t}}$ and including a primordial component in all the $B$-mode auto spectra.}. Crucially, our result thus predicts a suppression of the primordial signal whenever it is present, as first noted in ref.~\cite{ref:teng_11}. This effect is also hinted at in ref.~\cite{ref:namikawa_17}, though the fact that their simulations have $r=0$ prevents them from making a quantitative statement.

\begin{figure}
        \centering
        \includegraphics[width=\textwidth]{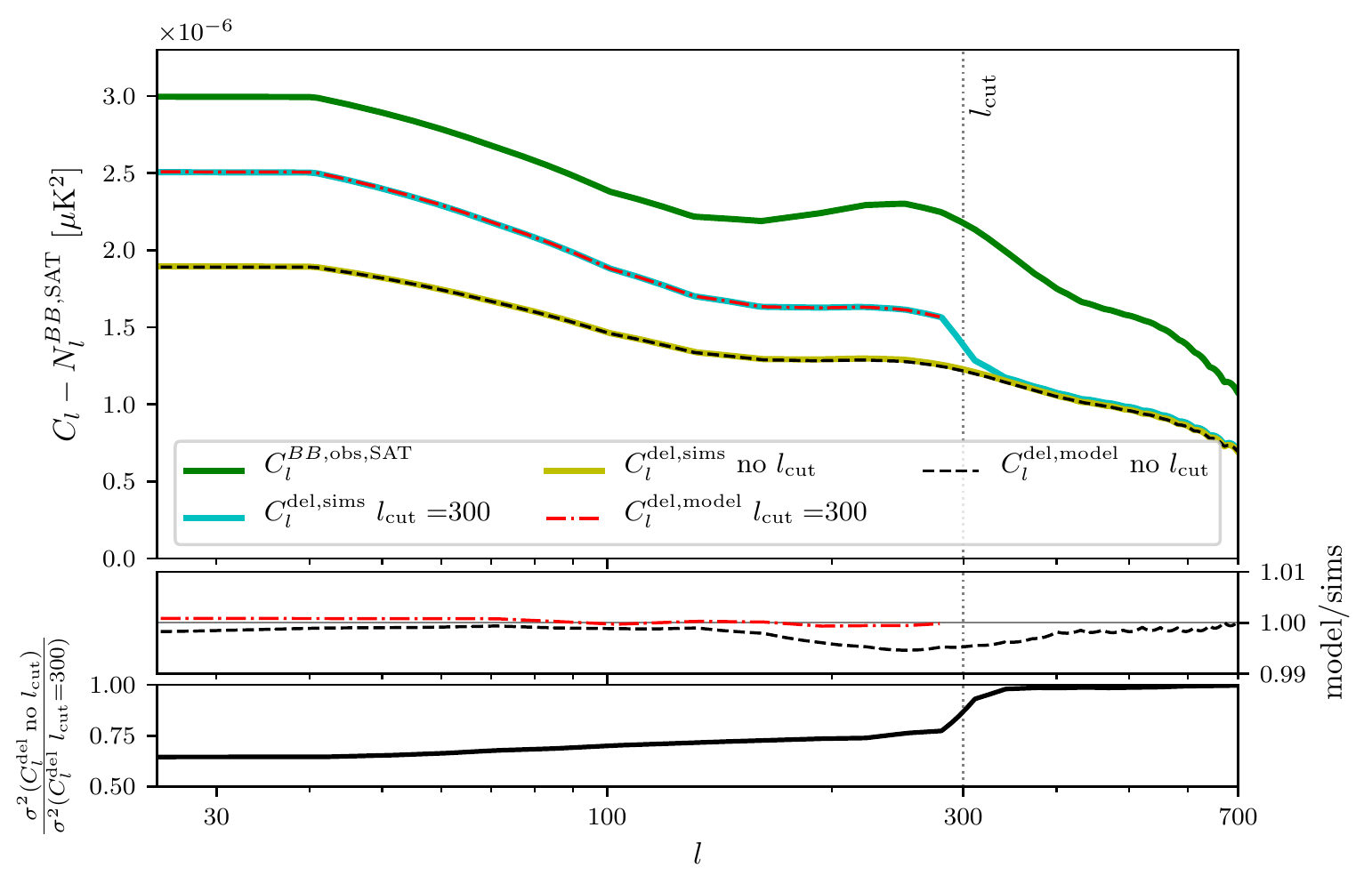}
        \caption{\textit{Upper panel}: noise-subtracted simulated power spectra of the observed $B$-mode (green), biased residual after delensing with $ \hat{\phi}^{EB} $ (yellow) and residual after delensing with a reconstruction cutoff at $l_{\mathrm{cut}}=300$ (cyan). The shaded regions represent the variance of each of the delensed spectra, obtained from simulations. Notice that the biased spectrum (yellow) has less variance than its unbiased counterpart (cyan). Also plotted are noise-subtracted theoretical curves modelling delensed spectra with (red, dot-dashed) and without (black, dashed) an $l_{\mathrm{cut}}$. \textit{Middle Panel}: ratio of model-to-simulated delensed power spectrum amplitude for the biased (black, dashed) and unbiased (red, dot-dashed) cases. \textit{Lower panel}: ratio of delensed power spectrum variance in the case where no $l_{\mathrm{cut}}$ is imposed to that where $l_{\mathrm{cut}}=300$, obtained from simulations. All curves are for an experiment with characteristics similar to the Simons Observatory, as described in the text.}
        \label{fig:all_spectra}
    \end{figure}

    In order to assess the validity of the model in eq.~\eqref{eqn:biased_delensed_spectrum_model}, we compare it to the output of applying the reconstruction and delensing pipelines to simulations (detailed in appendix~\ref{appendix:sims}) of an experiment with characteristics similar to those of SO~\cite{ref:so_science_paper}. Specifically, the LAT survey has angular resolution $\theta_{\text{FWHM}}=1.5\,\text{armin}$ and temperature noise level $\Delta_T = 6\,\mu\text{K\,arcmin}$ (with the polarisation level $\Delta_P = \sqrt{2} \Delta_T$), as in figure~\ref{fig:reconstruction_noise_levels}. It is the simulated data from this telescope which we use for the purpose of lensing reconstruction, with $l_{\mathrm{max}}=3000$. The SAT survey has $\theta_{\text{FWMH}} = 17\,\text{arcmin}$, $\Delta_T = 2 \,\mu\text{K\,arcmin}$ and $\Delta_P = \sqrt{2}\Delta_T$. Where needed for calculations of the power spectrum variance, we assume the SAT survey covers approximately $2.8\,\%$ of the sky (and is fully contained within the wider LAT survey). The simulations have tensor-to-scalar ratio $r=0.01$. The results are shown in figure~\ref{fig:all_spectra}.
    The biased delensed power spectrum (yellow curve in the top panel) is seen to be in excellent agreement with eq.~\eqref{eqn:biased_delensed_spectrum_model}, shown as the dashed black curve. The differences are smaller than $0.5\,\%$ on all scales. For lower noise levels, the level of agreement is expected to worsen as bias terms neglected by our model play an increasingly significant role. Since the most relevant of those terms (the contribution to the six-point function from the first term on the right of eq.~\ref{eq:fourptbiggest}) makes a positive contribution to the power, eq.~\eqref{eqn:biased_delensed_spectrum_model} is 
    expected to underestimate the amplitude of the biased delensed power spectrum. Indeed, we simulate a single-telescope experiment with $\theta_{\mathrm{FWHM}}=6\,\mathrm{arcmin}$ and $\Delta_{T}=3\,\mu \mathrm{K\,arcmin}$ and find that the modelled biased spectrum is around $2$--$3\,\%$ lower than its simulated counterpart.

    The bias in the delensed spectrum is around 80\,\% of the signal power, $C_l^{BB,t}+C_l^{BB,\text{res}}$, in the unbiased delensed spectrum for the configuration in figure~\ref{fig:all_spectra} (see discussion below). This bias would need to be modelled or otherwise mitigated in a likelihood analysis if inferences on the tensor-to-scalar ratio are not to be biased. One simple way to remove the bias in the delensed spectrum on the scales relevant for searches for primordial $B$-modes was noted by ref.~\cite{ref:teng_11}:
    exclude from the reconstruction any $B$-modes that overlap in scales with the $B$-modes we wish to delens. This follows from the local character of the bias, whose origin is given in eq.~\eqref{eqn:originoflocalbias},
    meaning that it will be avoided at a given multipole as long as \textit{that} multipole is removed from the
    $B$-field used for the lensing reconstruction. For the purpose of primordial $B$-mode searches, we shall remove reconstruction $B$-modes on the largest angular scales $l\leq l_{\mathrm{cut}}$. In this case, eq.~\eqref{eqn:originoflocalbias} becomes
    \begin{equation}
    \langle B^{\temp}(\bl) \rangle_{E^{\obs,\lat}} = B^{\obs,\lat}(\bl) D_l \Theta(l - l_{\text{cut}})\, ,  
    \end{equation}
    where $\Theta(l)$ is the Heaviside function, and so the bias terms in the delensed power spectrum vanish for $l\leq l_{\text{cut}}$. Note that the normalisation and noise of the $EB$ quadratic estimator are changed, albeit by a small amount, on all scales by excising the input $B$-modes at certain scales, and $D_l$ (for $l > l_{\text{cut}}$) needs to be calculated with the modified normalisation and Wiener filter; see eq.~\eqref{eqn:D_l}. Equivalently, we can think of effectively setting to infinity the noise on the $B$-modes used \emph{in the reconstruction} on the scales that we wish to exclude. In this case, $D_l$ goes to zero for $l\leq l_{\mathrm{cut}}$, and both eq.~\eqref{eqn:biased_delensed_spectrum_model} and~\eqref{eqn:biased_delensed_spectrum_model_single_tel} reduce to the unbiased spectrum of eq.~\eqref{eqn:unbiased_del_ps}. Averting the bias in this way comes at the cost of a lower signal-to-noise lensing reconstruction as information is being discarded, although the degradation is expected to be minor since most of the lensing information is obtained from smaller scales of the polarisation fields. In order to take this effect into account when computing the delensed power spectrum below the cutoff using eq.~\eqref{eqn:naive_clbbres}, we must make sure to calculate $C_l^{W}$ using a Wiener filter that includes the higher reconstruction noise. When this modification is included, the simple unbiased model agrees very well with the simulated spectrum, as shown in figure~\ref{fig:all_spectra} for the case of $l_{\text{cut}}=300$,
    with differences between the two less than $0.2\,\%$ for an SO-like experiment. Imposing a cutoff at $l_{\mathrm{cut}}=300$ leads to a lower signal-to-noise lensing reconstruction and a small decrease in $C_l^{W}$ which, on large angular scales, results in an increase of the residual lensing power spectrum and a degradation of delensing efficiency by approximately $2\,\%$.
    Several techniques have been suggested to reduce this small degradation, including realisation-dependent methods \cite{ref:namikawa_17} and an upgrade of the cutoff approach which involves splitting the whole multipole space into ``notches'' within which the bias can be avoided by excluding modes common to the notch and the reconstruction~\cite{ref:teng_11,ref:sehgal_17}.
    \begin{figure}
        \centering
        \includegraphics[width=0.7\textwidth]{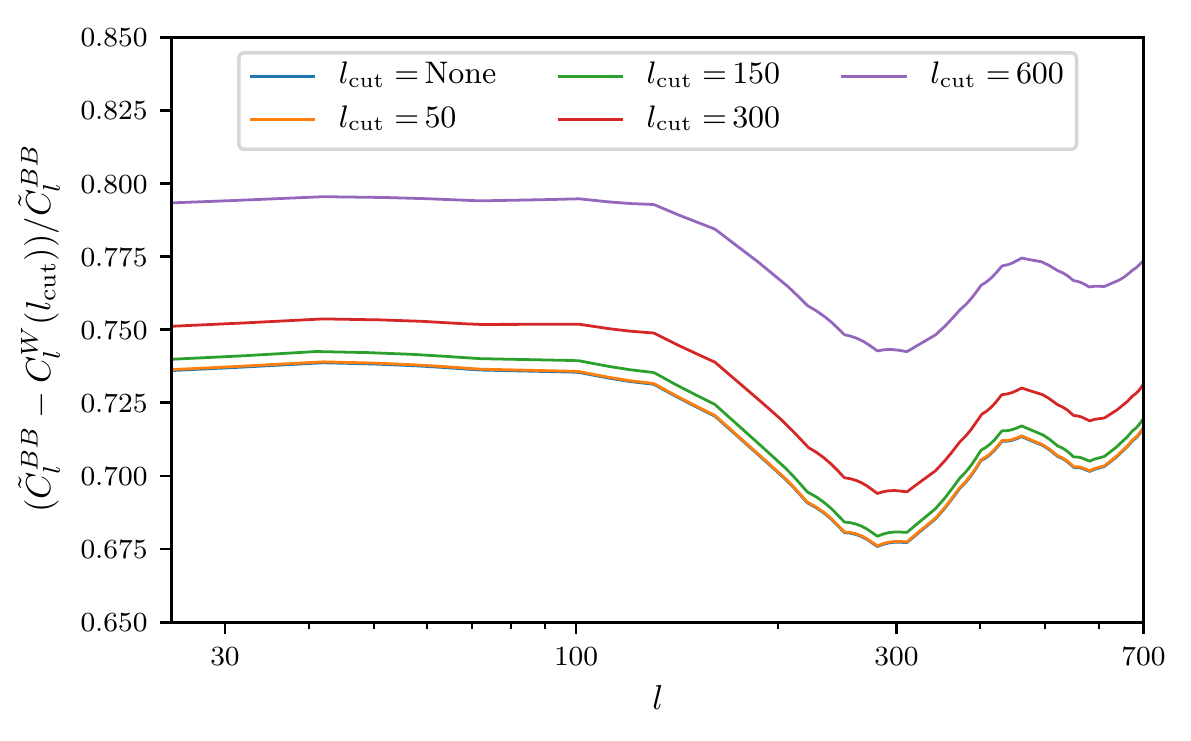}
        \caption{Remaining fraction of lensing $B$-mode power after delensing with an $EB$ quadratic estimator for different choices of the cutoff, for an experiment with LAT specifications as in figure~\ref{fig:reconstruction_noise_levels}.
    Note that the curves with $l_{\mathrm{cut}}=50$ and no $l_{\mathrm{cut}}$ overlap visually.}
        \label{fig:delensing_performance_vs_cutoff}
    \end{figure}

    Although the bias from internal delensing can be easily removed on large scales with the techniques just discussed, an alternative is to model explicitly the biased delensed spectrum as part of any subsequent likelihood analysis. Figure~\ref{fig:all_spectra} suggests it is worth exploring this approach further since the bias not only suppresses the delensed $B$-mode spectrum beyond what can be attributed to the (partial) removal of lensing effects, but also reduces its variance. Indeed, in figure~\ref{fig:all_spectra} the variance of the biased delensed spectrum is roughly $65\,\%$ of that associated with the unbiased case (the latter with $l_{\text{cut}}=300$). If the power spectrum bias were independent of the primordial $B$-mode power, this reduction in variance would translate directly into improved constraints on the tensor-to-scalar ratio (assuming there is no significant difference in the non-Gaussian correlations between the power spectra at different multipoles; see section~\ref{sec:covariances}). However, part of the bias does depend on the primordial spectrum, suppressing the contribution of $C_l^{BB,t}$ in the biased delensed spectrum by a factor $(D_l-1)^2$ (see eq.~\ref{eqn:biased_delensed_spectrum_model}).

\begin{figure}
        \centering
        \includegraphics[width=0.7\textwidth]{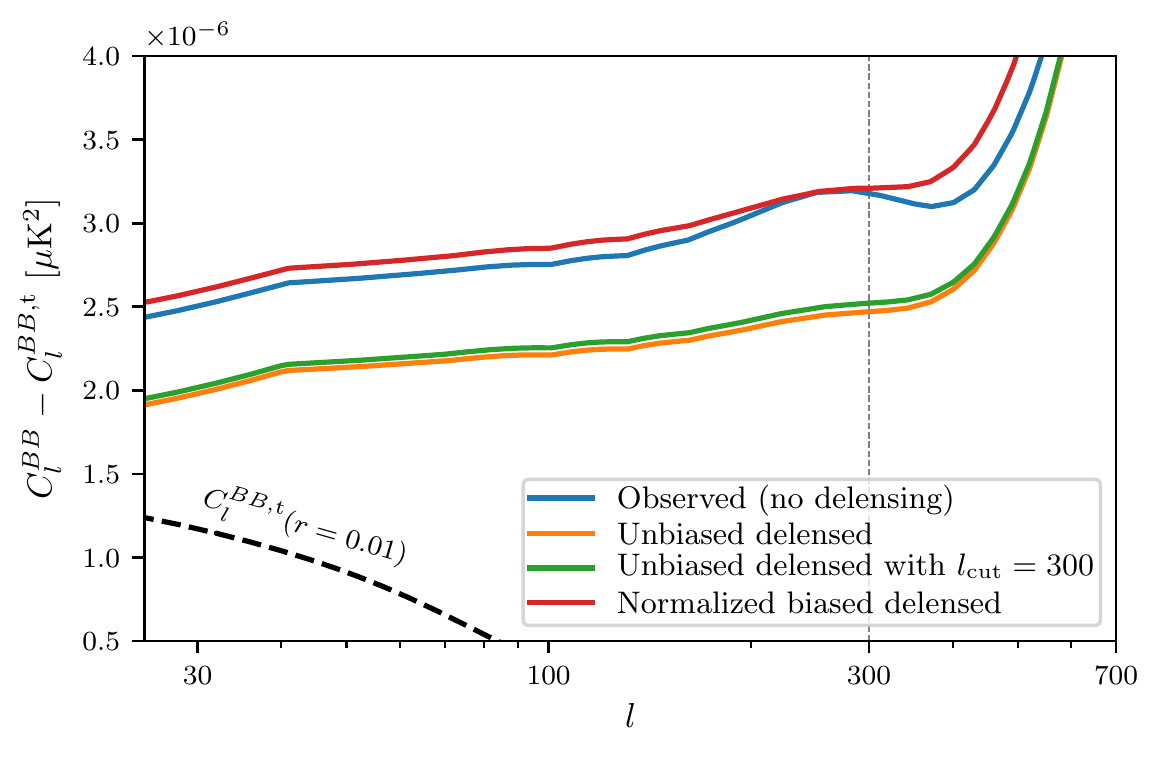}
        \caption{Effective noise power (i.e., the difference between the total delensed power and the primordial contribution) for different estimators from which a primordial $B$-mode signal (black dashed) is to be extracted. In the biased case, the spectrum is first renormalised so that the amplitude of the primordial power is the same as in the unbiased cases.
        The experimental set-up is the same as in figure~\ref{fig:all_spectra}.}
        \label{fig:renormalizing_delensed_spectra}
    \end{figure}

%
    The suppression of the primordial signal in the biased delensed spectrum acts as a multiplicative bias. We can remove this by renormalising by $(D_l-1)^{-2}$, in which case the contribution from $C_l^{BB,t}$ is the same as for unbiased delensing:
    \begin{equation}\label{eqn:renormalized_spectrum}
    \frac{C_l^{BB,\text{del}}}{(D_l-1)^2} = C_l^{BB,\text{del},\text{unbiased}} + \left( \frac{D_l}{D_l-1}\right)^2 \left[C_l^{W} + N_l^{BB,\lat} + N_l^{BB,\sat}\left(\frac{2}{D_l}-1\right) - \frac{2}{D_l} N_l^{X}\right],
    \end{equation}
    with $C_l^{BB,\text{del},\text{unbiased}}$ as defined in equation~\eqref{eqn:unbiased_del_ps}. The last three noise terms on the right-hand side vanish for the case of $B^{\obs,\lat} = B^{\obs,\sat}$, i.e., when the same maps are used in all parts of the analysis, and are positive when $B^{\obs,\lat}$ and $B^{\obs,\sat}$ have independent instrument noise ($N_l^X=0$) since $0 < D_l < 1$. On the other hand, the term proportional to $C_l^{W}$ is always positive. Crucially, this means that, in presence of the bias, the renormalised delensed power is generally larger than from the unbiased approach.\footnote{Mitigating the bias by excluding large-scale $B$-modes from the reconstruction will increase $C_l^{BB,\text{del},\text{unbiased}}$ a little since the reconstruction is noisier (figure~\ref{fig:delensing_performance_vs_cutoff}). For our instrument configuration, this increase is only around 2\,\% of $\tilde{C}_l^{BB}$ or 8\,\% of $C_l^W$ on large scales. By way of comparison,
    for the $N_l^X=0$ case, the instrument noise terms in square brackets in equation~\eqref{eqn:renormalized_spectrum} are about $25 C_l^W$ (taking $\tilde{C}_l^{BB}$ equivalent to $5\,\mu\text{K\,arcmin}$ white noise and $C_l^W = 0.25 \tilde{C}_l^{BB}$), and on multiplying by $D_l^2/(D_l-1)^2 \approx 0.045$, the additional power over $C_l^{BB,\text{del},\text{unbiased}}$
    is about $1.1 C_l^W$, which is worse than doing no delensing at all.} In other words,
    the primordial power is always suppressed by a larger fraction than the sources of (lensing and experimental) noise. This is illustrated in figure~\ref{fig:renormalizing_delensed_spectra} where we see that, for the experimental configuration adopted in figure~\ref{fig:all_spectra}, the signal-to-noise noise on a primordial component is lower when the bias is allowed to play a part and modelled than in the alternative approaches where the bias is avoided -- worse, even, than in the case of no delensing. In section~\ref{sec:likelihood}, we shall see that this indeed translates to errors on estimates of $r$ from a maximum-likelihood inference pipeline that inevitably increase (in a statistical sense) whenever the bias is modelled instead of avoided. For noise levels lower than those considered here, we expect the approximate expression for the biased delensed spectrum, equation~\eqref{eqn:biased_delensed_spectrum_model}, to underestimate the true power, as noted above, with the dominant correction coming from the first term on the right of eq.~\eqref{eq:fourptbiggest}. In this term, $B$-modes only enter through a connected four-point function of the form $\langle B E E B \rangle_c$, which receives no contribution from tensor modes (ignoring lensing for these). It follows that the additional term adds power but does not couple to $C_l^{BB,t}$, and so further reduces the signal-to-noise on the primordial signal.

    The renormalised spectrum of eq.~\eqref{eqn:renormalized_spectrum}
    is lower, all other things being equal, if the same maps are used for all parts of the analysis.
    The reason is that in this case the experimental noise in the reconstruction and in the $B$-modes to be delensed are correlated, allowing the disconnected couplings~\eqref{eqn:corrII} and~\eqref{eqn:corrtempxobsdisc} to suppress the experimental noise contribution to the delensed power spectrum by the same fraction as for the primordial signal. If, on the contrary, the noise is uncorrelated between the reconstruction and the $B$-modes to delens, it will not contribute to the disconnected term~\eqref{eqn:corrtempxobsdisc} and will consequently be suppressed by a smaller amount than the signal -- in fact, it will be amplified relative to the unbiased case by coupling~\eqref{eqn:corrII}.

    The degradation in signal-to-noise on the primordial $B$-mode power in the biased case can
    be understood by noticing that, in the case $N_l^{BB,\lat}=N_l^{BB,\sat} = N_l^X$, the disconnected couplings~\eqref{eqn:corrII} and~\eqref{eqn:corrtempxobsdisc} combine with the terms retained in the unbiased calculation to yield a suppression of the primordial power and experimental noise by an equal fraction, while for the lensing contribution there is an additional (positive) term, eq.~\eqref{eqn:corrIV}, which reinstates some lensing power while leaving the primordial signal unaffected. This last contribution arises from the connected coupling of the $B$-mode in one of the templates and the $EEB$ fields in the other template. This is not the only term that restores lensing power, since the partially-connected six-point function~\eqref{eqn:OIIA} coupling the lensing part of $\hat{{\phi}}$ across templates has a similar effect. However, the latter is already included in the unbiased calculation.
    
\section{\label{sec:covariances}Covariances of delensed power spectra}
\begin{figure*}
        \centering
        \includegraphics[width=\textwidth]{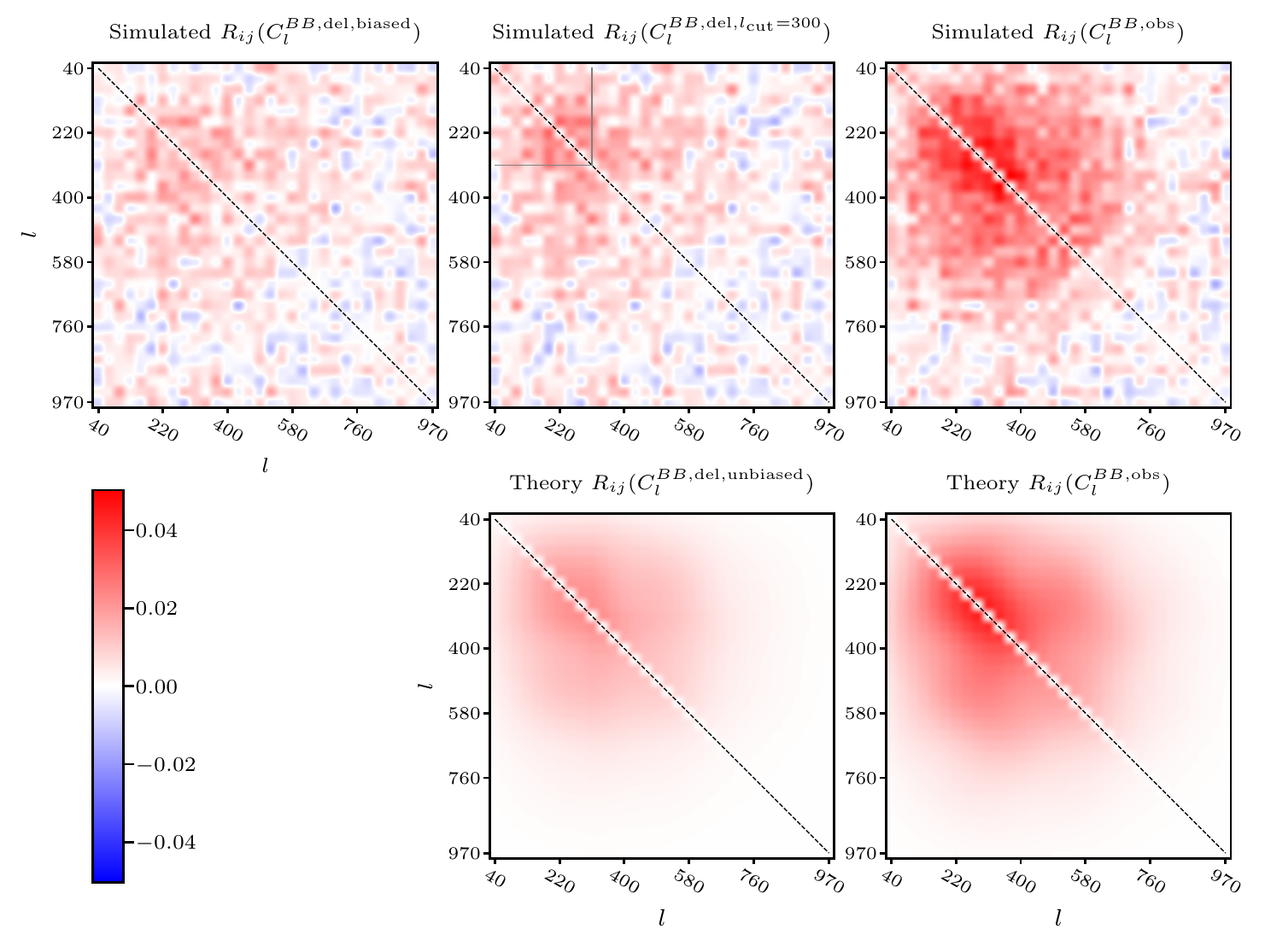}
        \caption{Correlation coefficient of $B$-mode bandpowers ($\Delta l =30$) for an experiment with characteristics similar to the Simons Observatory, as described in section~\ref{sec:biases}, and a SAT sky fraction of $2.8\,\%$ (with periodic boundary conditions assumed).
    The fiducial cosmology includes a primordial $B$-mode component with $r_{\mathrm{input}}=0.01$. The top row are obtained from 25\,000 simulations and the bottom row are analytic approximations evaluated by modifying \texttt{LensCov} \cite{ref:peloton_17}, as described in the text. \textit{Right column}: observed (lensed and noisy) bandpowers. \textit{Central column}: delensed bandpowers in the case where the bias is avoided by either introducing a cutoff at $l_{\mathrm{cut}}=300$ (top) -- thus avoiding the bias for $l<l_{\mathrm{cut}}$ -- or by working with a $\hat{\phi}$ that is independent of $B$ yet just as correlated with the actual $\phi$ as the reconstructed $\hat{\phi}^{EB}$ employed in the simulated analysis (bottom). \textit{Left column}: biased delensed bandpowers from simulations.}
        \label{fig:ps_corrcoeff_COMBINED_nsims_15000_display_1000}
    \end{figure*}

\begin{figure*}
    \centering
    \includegraphics[width=0.7\textwidth]{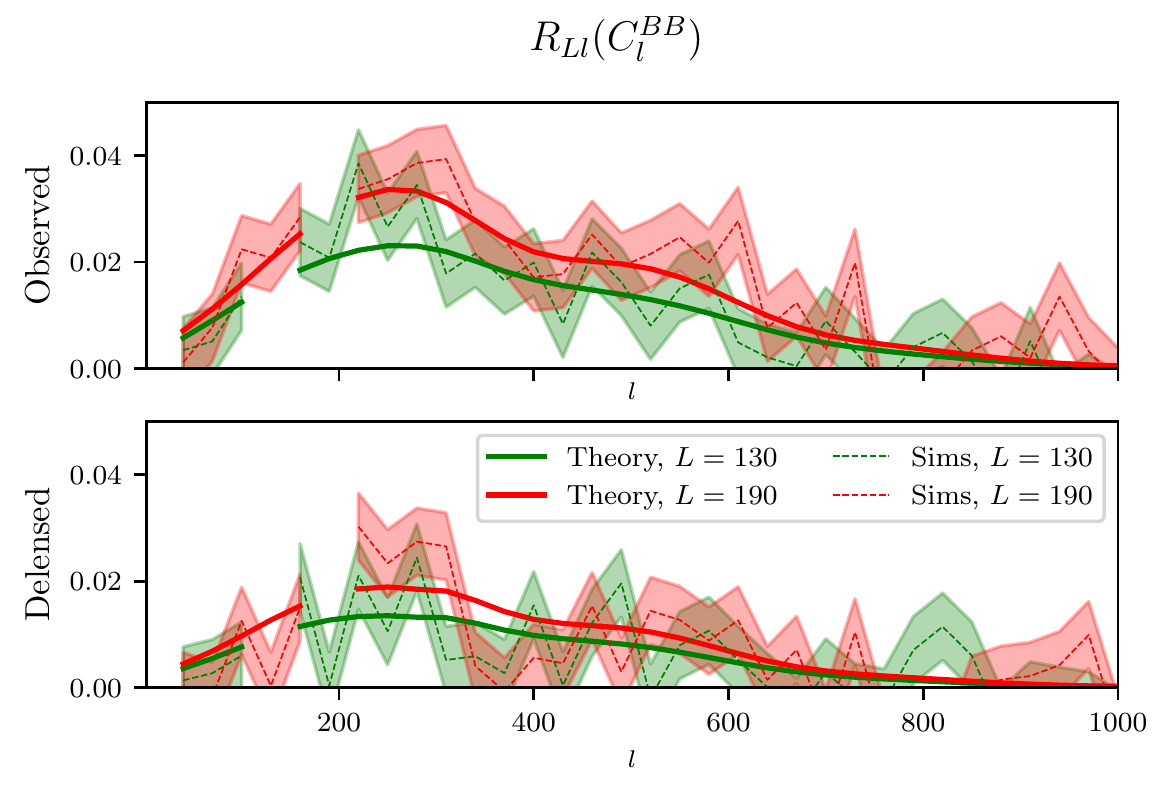}
    \caption{Rows of the correlation matrices of bandpowers of the observed (top) and delensed (bottom) $B$-modes shown in figure~\ref{fig:ps_corrcoeff_COMBINED_nsims_15000_display_1000}.
    Dashed curves correspond to simulated covariances, with the shaded region representing their approximate $1\,\sigma$ simulation error due to the finite number (25\,000) of simulations.
    In the delensed case, the simulations involve a reconstruction $l_{\mathrm{cut}}=300$ so that the delensed power is unbiased
    below $l<300$. Theory curves are shown as solid lines, taken from the bottom panels of figure~\ref{fig:ps_corrcoeff_COMBINED_nsims_15000_display_1000}.
    Gaps are shown where the correlation coefficient is one.}
    \label{fig:errorbars_rows_of_corrcoeff}
    \end{figure*}

    The precision with which we can isolate a primordial component from the observed (or delensed) $B$-mode power spectrum is ultimately determined by the covariance of the latter. The non-Gaussian covariance of lensed CMB spectra is well understood by now \cite{ref:smith_2004, ref:smith_challinor_2006, ref:benoit_levy_12, ref:peloton_17} and models exist that allow for its numerical evaluation. In this work, we use the publicly-available code \texttt{LensCov} \cite{ref:peloton_17} to compute the theoretical bandpower covariances for observations corresponding to our choice of cosmology and experimental parameters. We find good agreement between this theoretical computation and results from simulations after applying the appropriate binning and sky fraction corrections, as illustrated by figures~\ref{fig:ps_corrcoeff_COMBINED_nsims_15000_display_1000} and \ref{fig:errorbars_rows_of_corrcoeff}. For visualisation purposes, we plot the cross-correlation coefficient of bandpowers, defined as
    \begin{equation}\label{eqn:crosscorr_of_ps}
        R_{l_{1}l_{2}}(\tilde{C}^{BB}_l) \equiv \frac{\mathrm{Cov}(\tilde{C}^{BB}_{l_{1}}, \tilde{C}^{BB}_{l_{2}})}{\sqrt{\mathrm{Cov}(\tilde{C}^{BB}_{l_{1}}, \tilde{C}^{BB}_{l_{1}})\mathrm{Cov}(\tilde{C}^{BB}_{l_{2}}, \tilde{C}^{BB}_{l_{2}})}},
    \end{equation}
    where $\mathrm{Cov}(\tilde{C}^{BB}_{l_{1}}, \tilde{C}^{BB}_{l_{2}})$ is the covariance of the binned angular power spectrum
    \begin{equation}
        \mathrm{Cov}(\tilde{C}^{BB}_{l_{1}}, \tilde{C}^{BB}_{l_{2}}) = \langle\tilde{C}^{BB}_{l_{1}} \tilde{C}^{BB}_{l_{2}}\rangle - \langle\tilde{C}^{BB}_{l_{1}}\rangle\langle\tilde{C}^{BB}_{l_{2}}\rangle,
    \end{equation}
    and the angle brackets denote averaging over realisations of either the lensed or delensed CMB and noise.

    The delensed bandpower covariance is more complicated, but it can begin to be understood by studying the case where the lensing reconstruction is independent of the CMB. Reference~\cite{ref:namikawa_15} presents the following extension of the lensed $B$-mode power covariance to this delensed case under the assumptions that the $E$-modes used in the template are cosmic-variance limited on all relevant scales (i.e., for multipoles $l\lesssim 2000$) and that $\phi$ is Gaussian:
    \begin{align}\label{eqn:theory_signal_covariance}
        \mathrm{Cov}(\tilde{C}^{BB, \mathrm{del}}_{l_1}, \tilde{C}^{BB, \mathrm{del}}_{l_2}) &\approx \frac{2}{2l_1+1}\big(\tilde{C}_{l_1}^{BB, \mathrm{del}}\big)^2\delta_{l_{1}l_2}
        + \sum_l \bigg(\frac{\partial \tilde{C}_{l_1}^{BB, \mathrm{res}}}{\partial C_{l}^{EE}}  \frac{2}{2l+1}(C_l^{EE})^2\frac{\partial \tilde{C}_{l_2}^{BB, \mathrm{res}}}{\partial C_{l}^{EE}}\bigg)\nonumber\\
        \qquad &+ \sum_l \bigg(\frac{\partial \tilde{C}_{l_1}^{BB, \mathrm{res}}}{\partial C_{l}^{\phi\phi}}  \frac{2}{2l+1}(C_l^{\phi\phi})^2\frac{\partial \tilde{C}_{l_2}^{BB, \mathrm{res}}}{\partial C_{l}^{\phi\phi}}\bigg).
    \end{align}
    The assumption that $E$-modes are limited by cosmic variance on the relevant scales is not strictly true for the specifications adopted here. Although it is not difficult to generalise eq.~\eqref{eqn:theory_signal_covariance} to account for noise in the $E$-modes, ref.~\cite{ref:namikawa_15} finds that the covariance without $E$-mode noise still matches well that obtained from simulations of experiments with the noise levels of current and upcoming experiments. 

    We evaluate equation~\eqref{eqn:theory_signal_covariance} by modifying the \texttt{LensCov} code for the calculation of lensed $B$-mode covariances. Along with a modified Gaussian variance (the first term on the right of eq.~\ref{eqn:theory_signal_covariance}), we must also replace the derivative terms with those appearing above. We calculate these analytically from the leading-order expression of equation~\eqref{eqn:naive_clbbres}. The results are shown in figure~\ref{fig:ps_corrcoeff_COMBINED_nsims_15000_display_1000}. From this figure, and more acutely from figure~\ref{fig:errorbars_rows_of_corrcoeff}, it is clear that delensing reduces the non-Gaussian correlation between scales introduced by lensing~\cite{ref:smith_2004}, effectively increasing the number of independent pieces of information available. Figures~\ref{fig:ps_corrcoeff_COMBINED_nsims_15000_display_1000} and \ref{fig:errorbars_rows_of_corrcoeff} also suggest that the simple model of ref.~\cite{ref:namikawa_15} is consistent -- up to Monte-Carlo errors from the finite number (25\,000) of simulations\footnote{It can be shown (assuming that the errors on estimates of individual bandpowers are distributed as Gaussian random variables with appropriate correlations) that the fractional error on an element of the correlation matrix, estimated from $N$ simulations, is
    \begin{equation*}
        \frac{\sigma(R_{{ij}})}{|R_{ij}|} = \sqrt{\frac{1}{N}\left( 1 + \frac{1}{R^2_{ij}}\right)},
    \end{equation*}
    where $R_{ij}$ are the true correlation coefficients between bandpowers and the mean bandpowers are known in advance. In the more usual case where the variances are also estimated from the data, the relation is replaced with
    \begin{equation*}
        \frac{\sigma(R_{{ij}})}{|R_{ij}|} = \frac{1}{\sqrt{N}} \left| \frac{1}{R_{ij}} - R_{ij} \right|.
    \end{equation*}
    } -- with the simulated covariance of delensed $B$-mode bandpowers below the cutoff, where $\hat{\phi}$ is effectively independent of $B$.


    Finally, we study the bandpower covariance in the case where no $l_{\mathrm{cut}}$ is imposed and thus the delensed $B$-mode spectrum is biased. An analytic exploration of this case is beyond the scope of this paper, since the framework of appendix~\ref{appendix:calculation_of_eb_bias} suggests that the covariance receives contributions from 12-point correlators of lensed polarisation fields although some simplifications are likely possible. It can, however, be obtained from simulations, as shown in figure~\ref{fig:ps_corrcoeff_COMBINED_nsims_15000_display_1000}. Interestingly, at the noise levels considered here, the lower variance induced by the bias does not appear to be accompanied by the larger off-diagonal correlations between bandpowers seen by ref.~\cite{ref:namikawa_lss}. The most likely explanation is that the Gaussian experimental noise dominates the variance on the small angular scales where we would expect the bias to induce strong covariance between delensed bandpowers (cf.\ figure~11 of \cite{ref:namikawa_lss}). This suppresses the correlation and increases the fractional Monte-Carlo error to the point where the signal gets buried in the noise.

    In summary, we have seen that, for an experiment with characteristics similar to SO, the $EB$ lensing reconstruction bias brings about a reduction in the variance of the power spectrum of delensed $B$-modes while the extent of the correlations between bandpowers remains comparable to the unbiased case. If the primordial signal were not suppressed by the bias, analysis of the biased delensed bandpowers (with appropriate modelling of the bias) would lead to improved constraints on the tensor-to-scalar ratio. However, we have already seen in section~\ref{sec:biases}, that the bias acts multiplicatively on the primordial power, suppressing the signal more than the Gaussian variance. We thus expect constraints on $r$ to worsen in this case, which we now demonstrate through a simulated maximum-likelihood analysis.

\section{\label{sec:likelihood}Maximum-likelihood inference of $r$}
    Although the likelihood of CMB temperature and polarisation data can in principle be written down exactly, even accounting for the effects of lensing~\cite{ref:hirata_03_temperature,ref:carron_17_maximum,ref:millea_17}, in practical applications with realistic survey complexities, easily-computable summary statistics (e.g., angular power spectra) and their approximate sampling distributions are typically preferred (see~\cite{ref:cmb_likelihoods_review} for a review). For power spectra of approximately Gaussian fields, several such likelihood approximations involving the estimated power spectra and their covariance appear in the literature. We choose to work with the approximation developed in ref.~\cite{ref:hamimeche_2008}, and which was used by the BICEP/Keck collaboration to analyse their most recent data~\cite{ref:bicep2_18}. For our application here, it takes the form
    \begin{equation}\label{eqn:likelihood}
        -2\ln \mathcal{L}(r|\{\hat{C}^{BB}_i\}) = \sum_{ij}[g(\hat{C}^{BB}_i/C^{BB}_i)C^{BB,\fid}_i]\mathrm{Cov}^{-1}(C^{BB,\fid}_i, C^{BB,\fid}_j)[C^{BB,\fid}_jg(\hat{C}^{BB}_j/C^{BB}_j)],
    \end{equation}
    where $g(x)\equiv \text{sign}(x-1)\sqrt{2(x-\ln x -1)}$ and $\hat{C}^{BB}_i$ ($C^{BB}_i$) denotes the $i$th empirical (model) bandpower of the delensed $B$-mode spectrum. Only multipoles below $l_{\mathrm{max}}=280$ are used in the inferences that follow. An advantage of using this approximate likelihood is that the non-Gaussian character introduced by lensing can be accounted for through the fiducial delensed bandpower covariance, $\mathrm{Cov}(C^{BB,\fid}_i, C^{BB,\fid}_j)$, which only needs to be computed (either analytically or via simulations) and inverted once. As long as it matches the fiducial $C_l^{BB,\mathrm{fid}}$, the exact form of this fiducial covariance has a negligible impact on the resulting inferences. Indeed, we verify that there is no appreciable change in the inferences when we vary the fiducial level of primordial $B$-mode signal. Furthermore, ref.~\cite{ref:hamimeche_2008} shows that, even if the fiducial model deviates from the truth, the likelihood is still exact in the full-sky, isotropic limit with Gaussian fields. Hence, all analyses presented henceforth use fiducial bandpower amplitudes and covariances with the same $r_{\text{input}}$.

\begin{figure}
    \centering
    \includegraphics[width=0.7\textwidth]{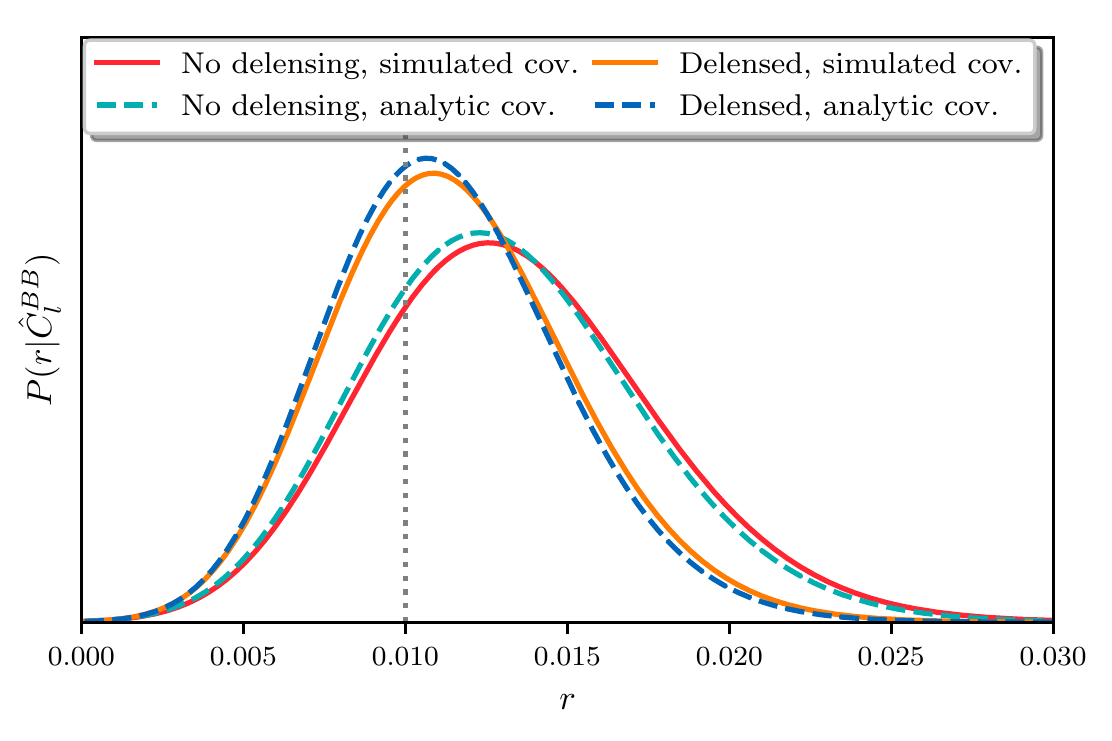}
    \caption{Typical likelihood functions for the tensor-to-scalar ratio $r$ for analyses of simulated data with and without (unbiased) delensing. In both cases, the use of simulated (solid lines) or analytic (dashed lines) power spectrum covariances is compared and found to be in good agreement. All curves are normalised to have the same value at the maximum-likelihood point. The input value $r_{\text{input}}=0.01$ is indicated with the dotted line. The simulations are for the experimental set-up described in section~\ref{sec:biases}.}
    \label{fig:prob_dist_of_r_modeling_vs_ignoring}
    \end{figure}

\begin{figure*}
    \centering
    \centering
    \includegraphics[width=\textwidth]{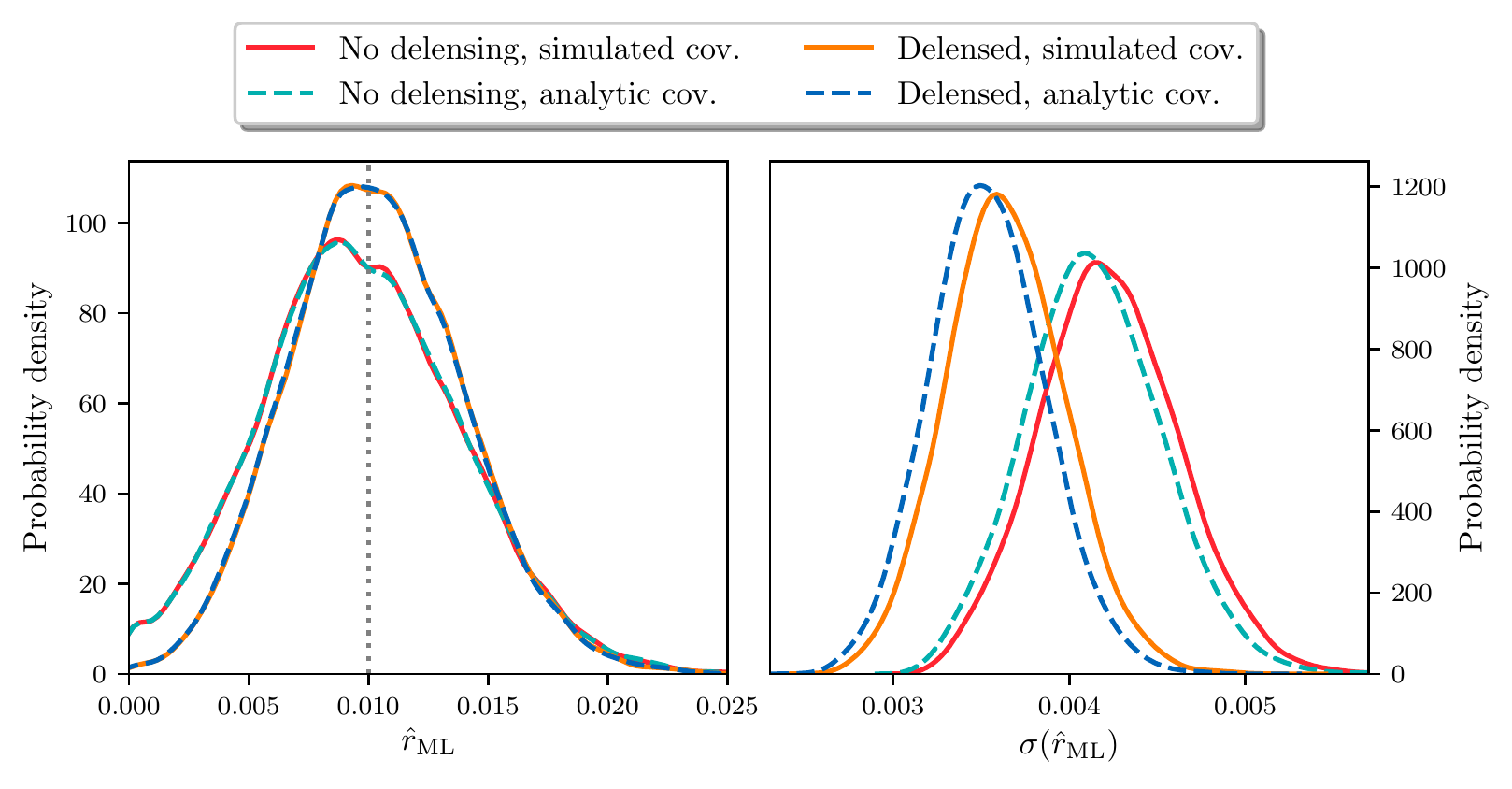}
    \caption{Distributions of $\hat{r}_{\mathrm{ML}}$ (left) and $\sigma(\hat{r}_{\mathrm{ML}})$ (right) across simulations comparing the use of analytic (dashed lines) or simulated (solid lines) power spectrum covariances in the cases of no delensing and unbiased delensing. The distributions have been estimated from 25\,000 simulated maximum-likelihood inferences, with a typical likelihood curve as shown in figure~\ref{fig:prob_dist_of_r_modeling_vs_ignoring}, using a Gaussian kernel density method with a bandwidth of $5\times10^{-4}$ for the left panel and $7\times10^{-5}$ for the right.}
    \label{fig:likelihood_dists_analyticvssimmed}
    \end{figure*}

    The question remains whether the bandpower covariances ought to be modelled or whether they could simply be obtained from simulations. In section~\ref{sec:covariances} we saw that, while it is possible to write a simple analytic model for the covariance of \textit{unbiased} delensed bandpowers, doing the same for the \textit{biased} case is more difficult. To address this question partially, we compare the distribution across simulations of the maximum-likelihood estimates for the tensor-to-scalar ratio $\hat{r}_{\mathrm{ML}}$ and their associated uncertainties $\sigma(\hat{r}_{\mathrm{ML}})$ -- derived from the second derivative of the log-likelihood function at the maximum-likelihood point -- in the cases where the bandpower covariance is obtained analytically (using \texttt{LensCov}) or from simulations. We do this for the cases of no delensing and unbiased delensing, the latter excluding $B$-modes below $l_{\mathrm{cut}}=300$ from the simulated fields used for reconstructing $\hat{\phi}^{EB}$. The simulations used in these comparisons, and for all results presented in this section, assume the experimental set-up described in section~\ref{sec:biases}. The likelihood curves for these two cases are shown in figure~\ref{fig:prob_dist_of_r_modeling_vs_ignoring} for a typical realisation, and the distribution of $\hat{r}_{\text{ML}}$ and $\sigma(\hat{r}_{\text{ML}})$ across 25\,000 simulations are given in figure~\ref{fig:likelihood_dists_analyticvssimmed}.
    We find only a slight degradation in the errors on $r$ when simulated bandpower covariances are employed, justifying our ensuing use of simulated matrices.


\begin{figure*}
    \centering
    \centering
    \includegraphics[width=\textwidth]{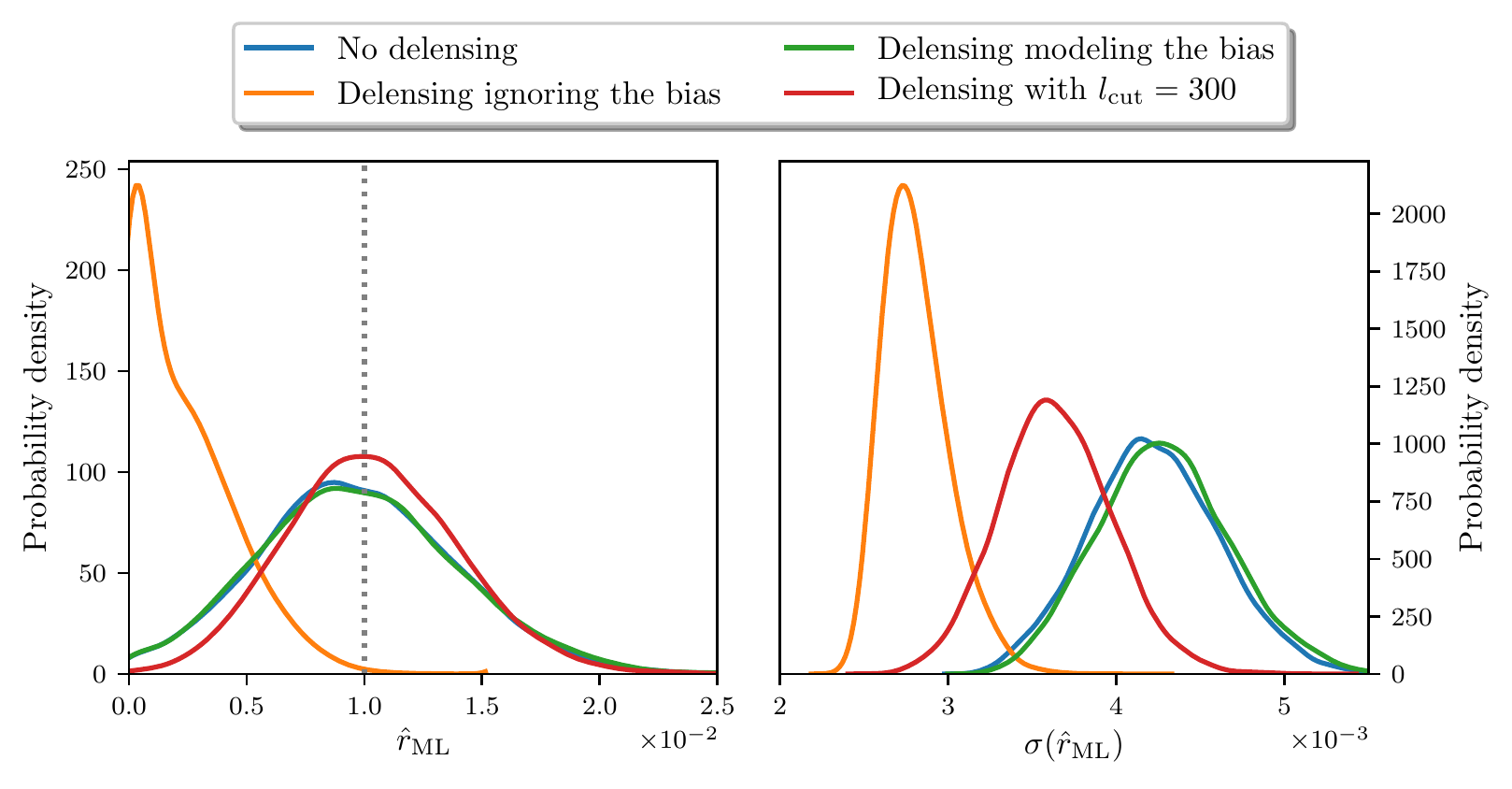}
    \caption{Distributions of $\hat{r}_{\text{ML}}$ (left) and $\sigma(\hat{r}_{\text{ML}})$ (right) across simulations for different delensing analyses. \textit{Blue}: no delensing. \textit{Orange}: biased delensing, based on a reconstructed $\hat{\phi}^{EB}$ that does not impose a low-$l$ $B$-mode cutoff, with no attempt to model the resulting power spectrum bias in the likelihood. This results in $\hat{r}_{\text{ML}}$ being biased low (and a significant number of simulated likelihoods that peak at $r=0$)
    and with artificially small errors. \textit{Green}: biased delensing, but with the power spectrum bias modelled as described in section~\ref{sec:biases}. This mitigates the bias in $r$ but, for our experimental set-up, has comparable performance to no delensing. \textit{Red}: a cutoff at $l_{\mathrm{cut}}=300$ is imposed on the $B$-modes used in the reconstruction, avoiding the bias to the delensed $B$-mode power spectrum below $l_{\mathrm{cut}}$. There is no bias in $r$ in this case, and delensing reduces $\sigma(\hat{r}_{\text{ML}})$ as intended. All analyses employ simulated bandpower covariances.}
    \label{fig:likelihood_dists}
    \end{figure*}

\begin{figure*}
    \centering
    \centering
    \includegraphics[width=\textwidth]{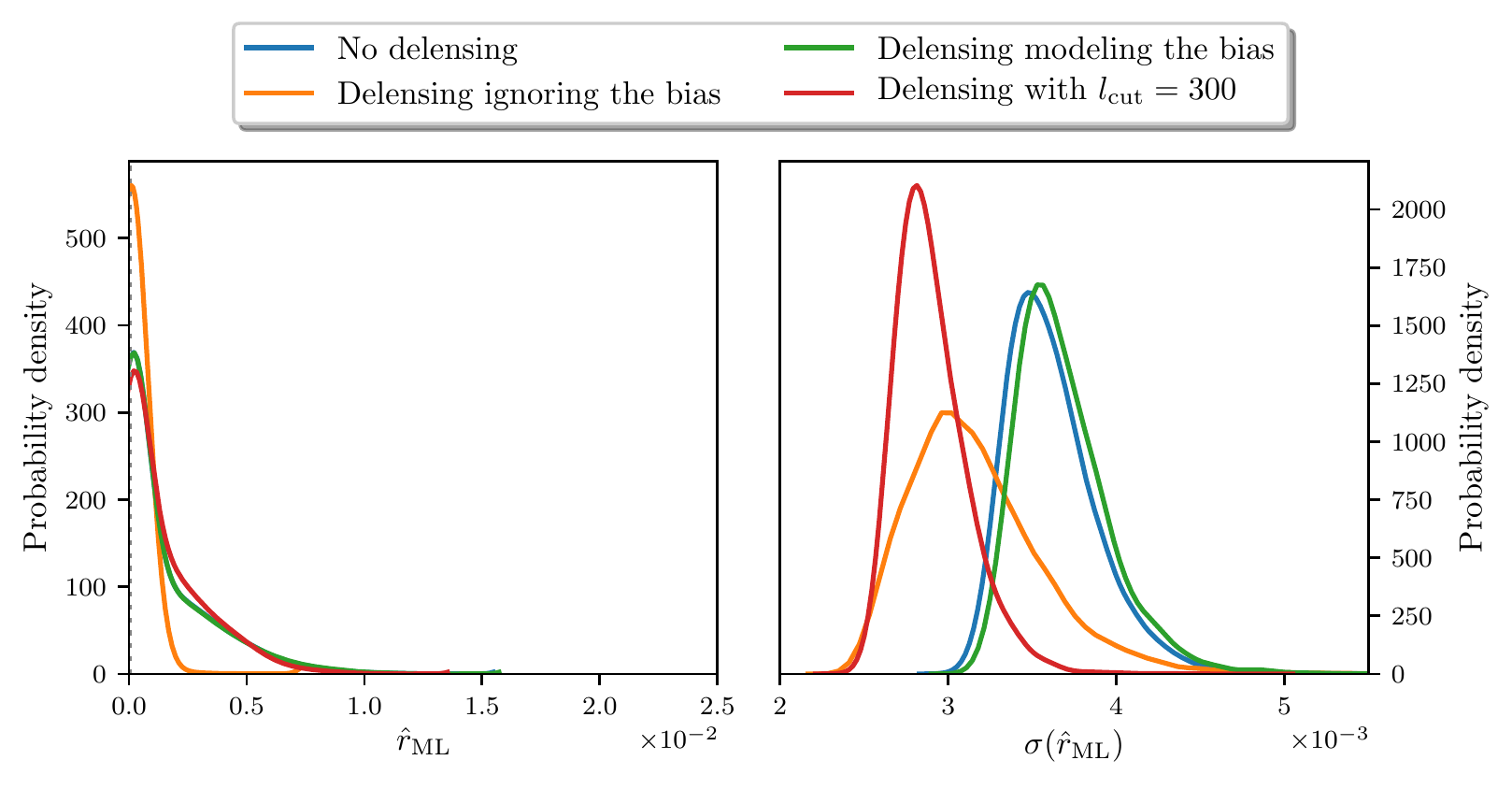}
    \caption{Same as figure~\ref{fig:likelihood_dists}, but in the null scenario of $r_{\mathrm{input}}=0$.}
    \label{fig:likelihood_dists_null}
    \end{figure*}

    We illustrate the need to account for (or mitigate) the bias in the delensed $B$-mode power spectrum when delensing with $\hat{\phi}^{EB}$ with the following naive analysis. For the empirical spectrum $\hat{C}_l^{BB}$ used in the likelihood of eq.~\eqref{eqn:likelihood}, we take the simulated biased delensed spectrum, but all other fiducial and model spectra and covariances in the likelihood are unbiased -- that is, calculated assuming that $\hat{\phi}$ is independent of the CMB but with the same correlation to $\phi$ as $\hat{\phi}^{EB}$. In figure~\ref{fig:likelihood_dists}, we verify that ignoring the suppression in the biased empirical spectrum leads to inferences on $r$ that are biased low relative to the input $r_{\mathrm{input}}=0.01$, with a significant number of likelihoods peaking (in the range $r\geq 0$) at $r=0$, and artificially small typical errors on $r$. This is also true in the null case where $r_{\mathrm{input}}=0$, as shown in figure~\ref{fig:likelihood_dists_null}.

    We have already discussed, in section~\ref{sec:biases}, several ways in which the bias can be taken into account: modelling the biased spectrum; renormalizing the biased spectrum to restore unit response to primordial power; or imposing a low-$l$ cutoff on the $B$ fields used for estimating $\hat{\phi}$. The first two approaches should yield equivalent results on $r$, but are expected to be very non-optimal since the bias reduces the primordial power by a larger fraction than the rest of the power (see figure~\ref{fig:renormalizing_delensed_spectra}). Mitigating the bias with a cut-off is preferred, although there is a small penalty due to the lower signal-to-noise lensing reconstruction in this case. In figure~\ref{fig:likelihood_dists}, we quantitatively compare these methods. For our experimental set-up and input value $r_{\text{input}}=0.01$, we see that modelling the bias inflates the errors on $r$ by typically $15\,\%$,\footnote{The exact figure depends on the experimental characteristics and $r_{\text{input}}$. We remind the reader that our likelihood results assume that lensing reconstruction is done exclusively by an $EB$ quadratic estimator and so underestimate the actual internal-delensing performance.} compared to when a cutoff at $l_{\mathrm{cut}}=300$ is imposed, translating to a slightly wider distribution of $\hat{r}_{\text{ML}}$ and a lower number of detections of non-zero $r$. The figure also shows that, for our set-up, modelling the bias degrades the sensitivity on $r$ to a level comparable to (or even slightly worse than) the case of no delensing, as suggested by our earlier results for the normalised power in this case (figure~\ref{fig:renormalizing_delensed_spectra}).

\section{\label{sec:obs_x_del_delensing} The $B^{\obs}(B^{\obs}-B^{\temp})$ estimator}
    In section~\ref{sec:biases}, we saw that the bias from internal delensing with an $EB$ estimator \emph{reduces} the signal-to-noise on a primordial contribution to the power spectrum of delensed $B$-modes. We attributed this to the coupling in eq.~\eqref{eqn:corrIV}, which arises in the cross-correlation of the $B$-mode lensing template with itself, and that
    restores some of the lensing power while leaving the primordial signal unaffected. This motivates the idea of considering an alternative power spectrum estimator of the form $B^{\obs}(B^{\obs}-B^{\temp})$, i.e., the cross-correlation of the observed $B$-modes with the delensed $B$-modes, $B^{\text{del}} = B^{\obs}-B^{\temp}$. This estimator does not involve the cross-correlation of the template with itself, and so avoids couplings such as that in eq.~\eqref{eqn:corrIV}. In the case of no $l_{\mathrm{cut}}$ in the lensing reconstruction (such that the $B$-modes used for reconstruction overlap with those to be delensed on all scales) and two separate surveys, LAT and SAT, the expected value of the cross-correlation is
    \begin{multline}\label{eqn:cross_estimator_biased_two_telescopes}
        \langle B^{\obs}(\bl)\left[B^{\obs}(\bl')-B^{\temp}(\bl')\right] \rangle_{\text{biased}} = (2\pi)^2\delta^{(2)}(\bl+\bl') \bigg[ (1-D_l)\left(C_l^{BB,\text{res}}+C_l^{BB,\mathrm{t}} + N_l^{BB,\mathrm{SAT}}\right) \\ 
        + D_l \left(N_l^{BB,\mathrm{SAT}} - N_l^{X}\right) - D_lC_l^W \bigg].
    \end{multline}
    In the case of a single survey, this reduces to
    \begin{multline}\label{eqn:cross_estimator_biased}
        \langle B^{\obs}(\bl)\left[B^{\obs}(\bl')-B^{\temp}(\bl')\right] \rangle_{\text{biased}} = (2\pi)^2\delta^{(2)}(\bl+\bl') \bigg[ (1-D_l)\left(C_l^{BB,\text{res}}+C_l^{BB,\mathrm{t}} + N_l^{BB}\right) \\ - D_l C_l^W\bigg].
    \end{multline}
    The last term on the right-hand side is always negative. Renormalising the cross-correlation by $(1-D_l)$ to have unit response to the primordial power, we see that its renormalised value is reduced below the power of the unbiased, delensed spectrum (equation~\ref{eqn:unbiased_del_ps}) by this negative term. While the non-primordial power in both of these spectra decreases monotonically with the experimental noise, the difference between the two is greatest when $C_l^W D_l/(1-D_l)$ peaks. This is the case when the experimental noise level in polarisation is approximately $3\sqrt{2}\,\mu\text{K\,arcmin}$, corresponding to a delensing efficiency of approximately $40\,\%$. For such an experimental configuration, the term $-D_l C_l^W$ reduces the cross-correlation in eq.~\eqref{eqn:cross_estimator_biased} by an amount equivalent on large scales to white noise with $\Delta_P \approx 2\,\mu\text{K\,arcmin}$ after normalising to have unit response to a primordial signal.

    It is instructive to compare the $B^{\obs}(B^{\obs}-B^{\temp})$ estimator above to another defined along similar lines, but which features no overlap between reconstruction $B$-modes and $B$-modes to be delensed (for example, by introducing an $l_{\mathrm{cut}}$ in the reconstruction, such that $\langle B^{\obs}(B^{\obs}-B^{\temp})\rangle $ is unbiased on scales $l<l_{\mathrm{cut}}$). This alternative estimator has expectation value
    \begin{equation}\label{eqn:cross_estimator_unbiased}
    \langle B^{\obs}(\bl)(B^{\obs}(\bl')-B^{\temp}(\bl')) \rangle_{\text{unbiased}} = (2\pi)^2\delta^{(2)}(\bl+\bl') \left( C_l^{BB,\text{res}}+C_l^{BB,\mathrm{t}} + N_l^{BB}\right),
    \end{equation}
    which is the same as the unbiased, delensed power spectrum of eq.~\eqref{eqn:unbiased_del_ps}. Comparing this to the biased estimator of eq.~\eqref{eqn:cross_estimator_biased}, we see that, after appropriate renormalisation, the expectation value of the latter is lower than the estimator in eq.~\eqref{eqn:cross_estimator_unbiased}, or even the standard power spectrum approach of eq.~\eqref{eqn:unbiased_del_ps}. Our analytic models for all these different estimators are found to be in excellent agreement with their values in simulations.

    In assessing the performance of the cross-correlation estimators for constraining the primordial signal, we remind the reader that, for a for a cross-correlation, the variance is not determined by the expected value (even in the limit of Gaussian fields).
    For a single survey, treating both $B^\obs$ and $B^{\text{del}}$ as Gaussian fields, the variances of the estimators at hand are
    \begin{align}
    \frac{1}{(1-D_l)^2}\text{Var}\left(B^{\obs}B^{\text{del}}\right)_\text{biased} &= \frac{1}{2l+1}\left[2 \left(\clbbdelunbias\right)^2  + \left(\frac{1-2D_l}{1-D_l}\right)^2 C_l^W \clbbdelunbias \right. \nonumber \\
    &{} \left.  + 2 \left(\frac{D_l}{1-D_l}\right)^2 \left( C_l^W \right)^2 \right] \, , \\
    \text{Var}\left(B^{\obs}B^{\text{del}}\right)_\text{unbiased} &= \frac{1}{2l+1}\left[2 \left(\clbbdelunbias\right)^2 +  C_l^W \clbbdelunbias \right] \, , \\
    \text{Var}\left(B^{\text{del}}B^{\text{del}}\right)_\text{unbiased} &= \frac{1}{2l+1}\left[2 \left(\clbbdelunbias\right)^2  \right] \, ,\label{eqn:gaussian_variance_delps}
    \end{align}
    where, recall, $\clbbdelunbias = C_l^{BB,\res} + C_l^{BB,t} + N_l^{BB}$. Clearly, the Gaussian variance is smallest for the usual $\left(B^{\text{del}}B^{\text{del}}\right)_\text{unbiased}$ estimator. Moreover, we will now see that the (co)-variance of both the cross-estimators is dominated by non-Gaussian effects, which further degrade their constraining power.

\begin{figure}
    \centering
    \includegraphics[width=0.7\textwidth]{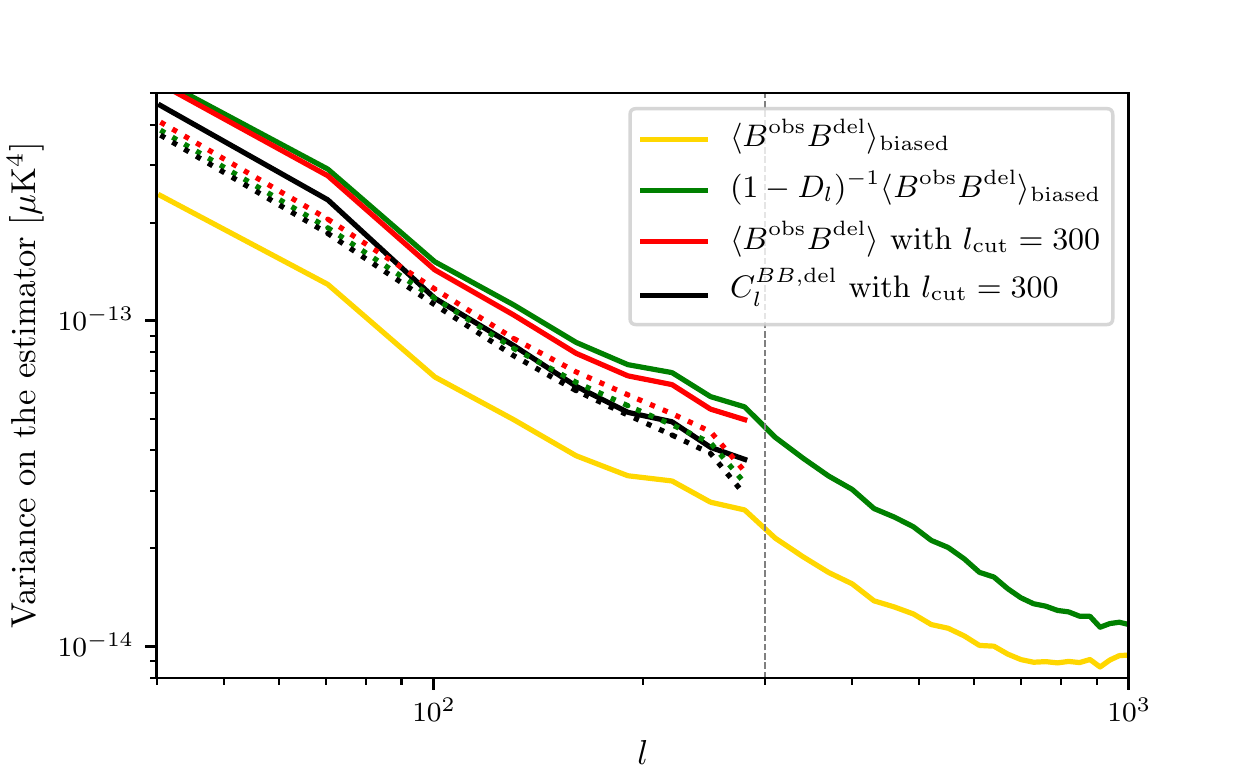}
    \caption{Variances of different spectral estimators from which primordial power is to be extracted, obtained from 25\,000 simulations of an experiment consisting of a single survey with $\Delta_{P}=3\sqrt{2}\,\mu \text{K\,arcmin}$, $\theta_{\mathrm{FWHM}}=6\, \mathrm{arcmin}$ and covering 2.8\,\% of the sky.
    \textit{Yellow}: the biased $(B^{\obs}B^{\text{del}})_{\text{biased}}$ estimator of eq.~\eqref{eqn:cross_estimator_biased}. \textit{Green}: the same for the case where $(B^{\obs}B^{\text{del}})_{\text{biased}}$ is normalised to have unit response to primordial $B$-mode power. \textit{Red}: the $(B^{\obs}B^{\text{del}})_{\text{unbiased}}$ estimator of eq.~\eqref{eqn:cross_estimator_biased} (only below $l_{\mathrm{cut}}=300$).  \textit{Black}: the auto-spectrum of $B^{\text{del}}$ in the case where the bias is averted below $l_{\mathrm{cut}}=300$. Dotted lines show the Gaussian variance expected for each estimator, calculated analytically.}
    \label{fig:obs_x_temp_variance}
    \end{figure}

\begin{figure}
        \centering
        \includegraphics[width=\textwidth]{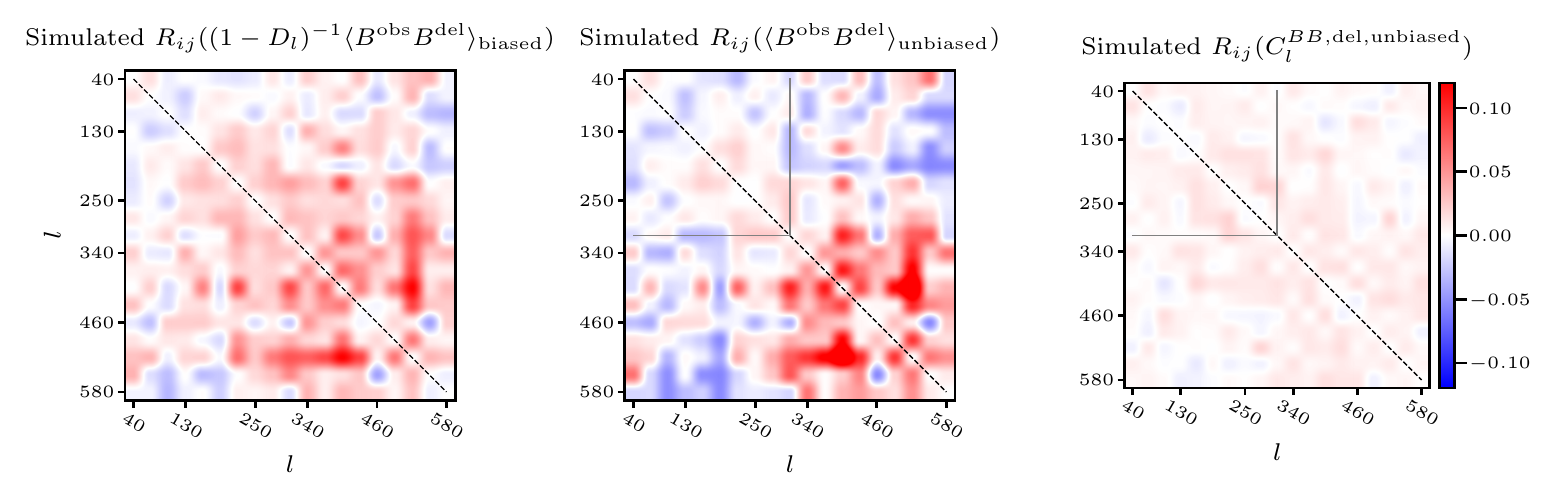}
        \caption{Cross-correlation coefficients of different spectral estimators for the experimental set-up in figure~\ref{fig:obs_x_temp_variance}, estimated from 25\,000 simulations. \textit{Right}: delensed $B$-mode bandpowers in the case where the bias is avoided below $l<l_{\mathrm{cut}}$ by introducing a cutoff at $l_{\mathrm{cut}}=300$. \textit{Centre}: $(B^{\obs}B^{\text{del}})_{\text{unbiased}}$ estimator of eq.~\eqref{eqn:cross_estimator_unbiased}, unbiased below $l_{\mathrm{cut}}=300$. \textit{Left}: $(B^{\obs}B^{\text{del}})_{\text{biased}}$ estimator of eq.~\eqref{eqn:cross_estimator_biased}, renormalised to have unit response to any primordial power. All estimators are binned with $\Delta l =30$.}
        \label{fig:obs_x_del_crosscorrs}
    \end{figure}

    In figure~\ref{fig:obs_x_temp_variance}, we use simulations of an experiment with $\Delta_{P}=3\sqrt{2}\,\mu\mathrm{K\,arcmin}$ and $\theta_{\mathrm{FWHM}}=6\,\mathrm{arcmin}$ to quantify the variance associated with the different estimators. Although the Gaussian variance of eq.~\eqref{eqn:gaussian_variance_delps} appears to capture correctly the simulated behaviour of the unbiased $\left(B^{\text{del}}B^{\text{del}}\right)$ estimator, the unbiased and renormalised $\left(B^{\text{obs}}B^{\text{del}}\right)$ estimators show variances well in excess of the Gaussian result. This can be attributed to the predominance of non-Gaussian contributions, which also manifest themselves in the strong correlations between bandpowers shown in figure~\ref{fig:obs_x_del_crosscorrs}. Intuitively, the reason why the non-Gaussian character is less acute in $\left(B^{\text{del}}B^{\text{del}}\right)$ than in $\left(B^{\text{obs}}B^{\text{del}}\right)$ is that the former involves two delensed fields and delensing is known to mitigate the correlations between scales introduced by lensing.

    Given these considerations, we believe that
    the estimator of eq.~\eqref{eqn:cross_estimator_biased} should lead to inferior constraints on $r$ compared to an analysis involving the usual power spectrum estimator of eq.~\eqref{eqn:unbiased_del_ps}. A more careful demonstration of this
    would require working within a maximum-likelihood inference framework, as in section~\ref{sec:likelihood}. Unfortunately, this is not straightforward since standard likelihood approximations are not applicable for a single cross-spectrum.
    These likelihood approximations are very convenient in the presence of real-world effects such as masking, and for including non-Gaussian power spectrum covariances. Extending these to a single cross-spectrum would require further development.

\section{\label{sec:discussion}Discussion}
    Searches for primordial $B$-modes in CMB data already necessarily require the removal of the contamination from gravitational lensing. In the near future, experimental characteristics will be such that the $EB$ quadratic estimator is expected to provide a large fraction of the signal-to-noise on reconstructions of the lensing potential, which is used to perform the delensing procedure. It is well known that any overlap in modes between the $B$-field to be delensed and the $B$-field from which the reconstruction is derived biases the amplitude and variance of the delensed power spectrum \cite{ref:teng_11,ref:namikawa_17}.

    In this paper, we have refined the modelling of this bias, based on a cumulant expansion and identification of the dominant terms, and used simulations to verify its efficacy. Our model is consistent with others in the literature~\cite{ref:namikawa_nagata_14, ref:namikawa_17} that are valid in the absence of a primordial (gravitational wave) signal, but extends our understanding to include the primordial component. The power spectrum bias acts multiplicatively on the primordial power, suppressing its amplitude. The bias also suppresses lensing and instrument noise power but, crucially, generally less so than for the primordial signal. This leads to a loss of constraining power for primordial $B$-modes, and this becomes more acute whenever separate surveys (i.e., with independent instrument noise) are used for lensing reconstruction and the study of $B$-modes on large angular scales.

    Fortunately, the bias is completely local in multipole space in the sense that bias to the delensed $B$-mode power at multipole $l$ originates from $B$-modes at that same multipole being used in the lensing reconstruction. The bias can therefore
    be avoided by eliminating overlapping modes in the $B$-field employed in the reconstruction, and within the context of searches for primordial gravitational waves this can be done with little loss of performance. We have shown that, for any reasonable $l_{\mathrm{cut}}$, such elimination of overlapping modes is actually preferable over an approach where the bias is modelled, as it yields improved constraining power. These findings are verified by simulating the reconstruction and delensing procedures and performing maximum-likelihood inferences of the primordial power on the resulting delensed spectra.

    Our model for the biased, delensed $B$-mode power spectrum is in excellent agreement with simulations for experimental noise levels comparable or inferior to those of the Simons Observatory, but we note that the agreement worsens for lower noise levels as new terms that we ignore become relevant. In this limit of low noise, we expect our model to over-estimate the signal-to-noise so that our claim that retaining, but modelling, the bias hinders efforts to constrain the primordial signal
    continues to hold. We therefore recommend that any attempt to delens large-scale $B$-modes internally building on lensing information gleaned from the $B$-modes themselves -- be it via quadratic estimators or likelihood-based techniques -- feature the excision of large-scale $B$-modes from the lensing inference.

    Finally, we have considered alternative cross-correlation estimators for the primordial power involving the observed and delensed $B$-modes, with a view to removing some couplings in the biased spectrum that degrade the signal-to-noise on $r$. We provided analytic models for the expectation values of these cross-correlations and verified these against simulations. However, we showed that ultimately these estimators have lower signal-to-noise to primordial power than the auto-power spectrum of the (unbiased) delensed fields, due to their cross-correlation character and their increased non-Gaussian covariance between multipoles.

    
\section*{Acknowledgements}
    ABL acknowledges support from an Isaac Newton Studentship at the University of Cambridge and from the Science and Technology Facilities Council (STFC). AC acknowledges support from the STFC (grant numbers ST/N000927/1 and ST/S000623/1). JC acknowledges support from a SNSF Eccellenza Professorial Fellowship (No. 186879) and from the European Research Council under the European Union's Seventh Framework Programme (FP/2007-2013) / ERC Grant Agreement No. [616170]. We thank Toshiya Namikawa and Blake Sherwin for useful discussions.

\appendix
\section{Internally-delensed $B$-mode power spectrum}\label{appendixa}

    In this appendix, we give details of the calculation of the $B$-mode power spectrum after internal delensing, as given in eq.~\eqref{eqn:biased_delensed_spectrum_model} in the main text. Specifically, we evaluate the four- and six-point functions of the observed CMB that appear in the cross-correlation of the lensing $B$-mode template with the observed $B$-modes, eq.~\eqref{eqn:temp_x_obs}, and the power spectrum of the template, eq.~\eqref{eqn:temp_x_temp}, respectively. 
    
    It is convenient to expand these four- and six-point functions in terms of connected $n$-point functions. In the absence of lensing, the CMB would be Gaussian and all connected $n$-point functions with $n>2$ would vanish. As the CMB fields are zero mean, the evaluation of $n$-point functions would reduce to products of two-point functions. However, in section~\ref{sec:covariances} we saw that lensing distorts the Gaussian primordial statistics, introducing significant non-Gaussianities in the form of non-vanishing, higher-order connected $n$-point functions, which we shall also refer to as ``connected correlators''. In particular, the connected four-point function, or trispectrum, induced by lensing lies at the heart of any effort to infer $C_L^{\phi\phi}$ from the lensed CMB \cite{ref:hu_2001_trispectrum, ref:cooray_kesden_03, ref:kesden_cooray_kamionkowski_03, ref:hanson_11}. Note that Gaussian instrumental noise does not contribute to the connected correlators of the observed CMB fields. Moreover, if we ignore the impact of lensing on the (otherwise Gaussian) gravitational wave contributions to the CMB (which should be a good approximation given their power falls rapidly on intermediate and small scales), these contributions are independent of the lensed, scalar contribution and so do not contribute to the connected correlators either.

    We organise the expansion of the $n$-point functions with the aid of a graphical representation where fields are represented by nodes, drawing those that are observed by the LAT in red, and those observed by the SAT in green. Lines connecting $n$ nodes denote the connected $n$-point function of the associated fields, which can be evaluated perturbatively to the desired order in lensing. In order to preserve generality, we ensure that the case where a single telescope is used for both construction of the delensing template and the observation of the large-scale $B$-modes can be recovered by letting $N_l^{X}=N_l^{BB,\lat}=N_l^{BB,\sat}$, with $N_l^{X}=0$ otherwise.

    In this way, the four-point function appearing in eq.~\eqref{eqn:temp_x_obs} can be represented as
    \begin{equation}\label{eqn:temp_x_obs_diags}
    \langle E^{\obs,\,\lat}(\bl'_1) E^{\obs,\,\lat}(\bl''_1) B^{\obs,\,\lat}(\bl_1-\bl'_1-\bl''_1)B^{\obs,\,\sat}(\bl_2)\rangle = \corrtempxobsconn + \corrtempxobsdisc \,,
    \end{equation}
    where terms of the form $ \langle EB\rangle $ have vanished due to parity invariance. The $B$- and $E$-fields involved in the lens reconstruction are at the top and bottom of the diagrams, respectively, and the $E$-modes used further in the template are at the midpoint on the left. The two diagrams on the right of eq.~\eqref{eqn:temp_x_obs_diags} correspond to a trispectrum $ \langle E^{\text{obs}}E^{\text{obs}}B^{\text{obs}}B^{\text{obs}}\rangle _c$ and a product of two-point functions $ \langle E^{\text{obs}}E^{\text{obs}}\rangle\langle B^{\text{obs}}B^{\text{obs}}\rangle$. Similarly, the six-point function appearing in eq.~\eqref{eqn:temp_x_temp} can be decomposed into
    \begin{align}\label{eqn:temp_x_temp_diags}
    &\langle E^{\obs,\,\lat}(\bl'_1) E^{\obs,\,\lat}(\bl''_1) B^{\obs,\,\lat}(\bl_1-\bl'_1-\bl''_1)E^{\obs,\,\lat}(\bl'_2) E^{\obs,\,\lat}(\bl''_2)B^{\obs,\,\lat}(\bl_2-\bl'_2-\bl''_2)\rangle \nonumber\\
    & \quad = \corrOI +  2\times\,\corrVI +  \corrOII + \corrIII + \corrI \nonumber\\
    &\qquad + 2\times\,\corrIV + \corrII  + \corrforgottenI + \corrOIII \,.
    \end{align}
    %
    Diagrams shown with multiplicity factors of two correspond to two diagrams related by $\bl_1 \leftrightarrow \bl_2$ (and similarly for primed wavevectors). 
    
    \subsection{Contributions that appear in the unbiased delensed power spectrum}\label{appendix:std_calc_from_full}

    The standard calculation of eq.~\eqref{eqn:naive_clbbres} assumes that the statistical noise in the lensing reconstruction, $ \hat{\phi} $, that appears in $B^{\temp}$ is independent of the CMB fields. For this reason, it includes only some of the terms making up eqs.~\eqref{eqn:temp_x_obs} and \eqref{eqn:temp_x_temp} or, equivalently, some of the couplings represented by the diagrams of eqs.~\eqref{eqn:temp_x_obs_diags} and~\eqref{eqn:temp_x_temp_diags}. We calculate these here, before assessing the remaining terms in section~\ref{appendix:calculation_of_eb_bias}.
    
    We start with the cross-correlation of the template with the observed $B$-modes, i.e., $ -2\langle B^{\temp}(\bl_1) B^{\obs}(\bl_2)\rangle $. Only the part involving the connected four-point function is included in the standard calculation, and then only the ``primary coupling'' in which there is a contraction over the unlensed $E$-modes across the two legs of the quadratic estimator $\hat{\phi}$. This yields
    \begin{equation}\label{eqn:contrib_to_std_from_tempxobsconn}
    \begin{split}
    -2\times\corrtempxobsconn &\supset -2\,\wick{\langle \tilde{E}[\c1 E,\phi] \tilde{E}[\c2 E,\phi] \tilde{B}[\c2 E,\c3 \phi] \tilde{B}[\c1 E,\c3 \phi]\rangle}\\
    &\rightarrow -2\,(2\pi)^2 \delta^{(2)}(\bl_1+\bl_2) C_{l_1}^{W} ,
    \end{split}
    \end{equation}
    where $ C_{l}^{W} $ is defined in eq.~\eqref{eqn:def_clbbW}. Here, we have introduced the notation, for example, $\tilde{E}[E,\phi]$ to show the functional dependence of the field $\tilde{E}$ on the unlensed $E$-modes and $\phi$. Note that the dependence on the unlensed $E$-modes is linear, and where $\phi$ is uncontracted, the unlensed field $E$ is implied. 
    The one other possible trispectrum coupling is subdominant. Ultimately, our justification of this is the good agreement between our model for the delensed $B$-mode power spectrum and our simulation results. However, a plausibility argument can be made based on the volume of wavevector space that terms can accumulate in the integral in eq.~\eqref{eqn:temp_x_obs}, similar to the reasoning why the primary coupling should dominate over other ``$N^{(1)}$'' couplings in the auto-power spectrum in CMB lensing reconstruction~\cite{ref:hu_2001_trispectrum}. Generally, terms that couple together two or more of the weights [$W(\bl_a,\bl_b)$] in the integrand, i.e., the least factorisable terms, will be subdominant relative to other terms where the weights are uncoupled and some of the nested integrals can be separated, since the volume of wavevector space is reduced in the former case. In the specific case of eq.~\eqref{eqn:contrib_to_std_from_tempxobsconn}, the primary coupling produces the same pattern of weights as appears in the rest of the integrand in eq.~\eqref{eqn:temp_x_obs}, while for the other coupling there is no such factorisation.
    The primary coupling is responsible for the removal of the lensing signal in $B^{\obs}$ that correlates with $B^{\temp}$; that is, the actual delensing.
    
    We now consider contributions from the six-point function $ \langle B^{\temp}(\bl_1) B^{\temp}(\bl_2)\rangle $. The only fully-disconnected term that appears in the standard calculation is
    \begin{align}\label{eqn:OIII_notsimplified}
    \corrOIII \rightarrow (2\pi)^2 \delta^{(2)}(\bl_1+\bl_2)\int& \frac{d^2\bl'_1}{(2\pi)^2} W^2(\bl_1, \bl'_1)\left(\Welat_{l'_1}\right)^2 C_{l'_1}^{EE,\obs,\lat} \left(\Wp_{|\bl_1-\bl'_1|}\right)^2 \nonumber \\
    & \times \left(A^{EB}_{|\bl_1-\bl'_1|}\right)^2 \int \frac{d^2\bl''_1}{(2\pi)^2} W^2(\bl_1-\bl'_1-\bl''_1,-\bl''_1)
    \left(\tilde{C}_{l''_1}^{EE, \fid}\right)^2 \nonumber \\
    & \quad \times \frac{C_{l''_1}^{EE,\obs,\lat}C_{|\bl_1-\bl'_1-\bl''_1|}^{BB,\obs,\lat}}{\left(C_{l''_1}^{EE,\obs,\fid,\lat}
    C_{|\bl_1-\bl'_1-\bl''_1|}^{BB,\obs,\fid,\lat}\right)^2} .
    \end{align}
    Typically, the fiducial observational power spectra used to inverse-variance weight the CMB fields for lensing reconstruction, and in the Wiener filters, will be calibrated from the observed power so to an excellent approximation $C_{l}^{EE,\obs,\lat} \approx C_{l}^{EE,\obs,\fid,\lat}$, and similarly for the $B$-mode spectra. In this case, the integral over $\bl''_1$, combined with the (fiducial) normalisation $(A^{EB}_{|\bl_1-\bl'_1|})^2$, gives the Gaussian reconstruction noise $N^{(0)EB}_{|\bl_1-\bl'_1|}$ of the quadratic estimator. Finally, if we substitute $\Welat_{l} C_{l}^{EE, \obs,\lat}\approx C_{l}^{EE, \fid}$ in eq.~\eqref{eqn:OIII_notsimplified}, we obtain
    \begin{equation}\label{eqn:OIII}
    \corrOIII \rightarrow (2\pi)^2 \delta^{(2)}(\bl_1+\bl_2)\int \frac{d^2\bm{l}'_1}{(2\pi)^2} W^2(\bl_1, \bl'_1)\left[\Welat_{l'_1} C_{l'_1}^{EE, \fid}\right]\left[\left(\Wp_{|\bl_1-\bl'_1|}\right)^2 N_{|\bl_1-\bl'_1|}^{(0) EB}\right].
    \end{equation}

    The only other diagram that is retained in the standard calculation involves a Gaussian correlation of $E$-modes across templates multiplying a trispectrum made up of two quadratic estimators, $\langle\hat{\phi}\hat{\phi}\rangle_c$.
    This trispectrum has been studied in detail by ref.~\cite{ref:cooray_kesden_03}. To $\mathcal{O}(C_L^{\phi\phi})$, it 
    evaluates to a sum of the lensing power spectrum, from the primary coupling, and an $N^{(1)}$ term from the other couplings.
    Only the primary coupling is included in the standard calculation, and arises from contractions of the form 
    \begin{equation}\label{eqn:OIIA}
    \corrOII \, \supset \, \wick{\langle \hat{\phi}^{EB}[\c1 E,\tilde{B}[\c1 E, \c2 \phi]] \hat{\phi}^{EB}[\c3 E,\tilde{B}[\c3 E,\c2 \phi]]\rangle}. 
    \end{equation}
    However, we include $N^{(1)}$ here for completeness to find
    \begin{multline}\label{eqn:OIIA_allcontribs}
    \corrOII \,  \rightarrow \, (2\pi)^2 \delta^{(2)}(\bl_1+\bl_2) \int \frac{d^2\bm{l}'_1}{(2\pi)^2} W^2(\bl_1, \bl'_1)\left[\Welat_{l'_1} C_{l'_1}^{EE, \fid}\right] \\
    \times \left[\left(\Wp_{|\bl_1-\bl'_1|}\right)^2 \left(C_{|\bl_1-\bl'_1|}^{\phi\phi, \mathrm{fid}} + N_{|\bl_1-\bl'_1|}^{(1)EB}\right)\right],
    \end{multline}
    where we have assumed, again, that $\Welat_{l} C_{l}^{EE, \obs,\lat}\approx C_{l}^{EE, \fid}$.
    The explicit form of $N_{|\bl_1-\bl'|}^{(1) EB}$ is given in eq.~(57) of ref.~\cite{ref:cooray_kesden_03}. We find this contribution to be subdominant on the relevant scales to both \eqref{eqn:OIII} and to the contribution from the primary coupling of the trispectrum, which gives rise to $C_l^{\phi\phi}$ in the integrand of~\eqref{eqn:OIIA_allcontribs}.

    Combining eqs.~\eqref{eqn:OIII} and~\eqref{eqn:OIIA} gives
    \begin{align}
    \corrOIII + \corrOII&\rightarrow(2\pi)^2 \delta^{(2)}(\bl_1+\bl_2) \int \frac{d^2\bm{l}'_1}{(2\pi)^2}  W^2(\bl_1, \bl_1')\left[\Welat_{l_1'} C_{l_1'}^{EE, \fid}\right]\nonumber\\
    &\qquad \qquad \qquad \qquad \qquad\times \left[(\Wp_{|\bl_1-\bl_1'|})^2\left(C^{\phi\phi, \mathrm{fid}}_{|\bl_1-\bl_1'|} + N_{|\bl_1-\bl_1'|}^{(0) EB}+N_{|\bl_1-\bl_1'|}^{(1) EB}\right)\right]\nonumber\\
    &\approx (2\pi)^2 \delta^{(2)}(\bl_1+\bl_2) C_{l_1}^{W},
    \end{align}
    where in the last line we have neglected $N_{|\bl_1-\bl'|}^{(1) EB}$
    and assumed that $\Wp_l (C_l^{\phi\phi} + N_l^{(0)EB})\approx C_l^{\phi\phi}$. Together with eq.~\eqref{eqn:obs_x_obs} and eq.~\eqref{eqn:contrib_to_std_from_tempxobsconn}, we recover the standard result for the residual lensing power spectrum of eq.~\eqref{eqn:naive_clbbres}:
    \begin{equation}\label{eqn:std_calc}
    \langle B^{\mathrm{del}}(\bl_1)B^{\mathrm{del}}(\bl_2)\rangle = (2\pi)^2 \delta^{(2)}(\bl_1+\bl_2) \left(N_{l_1}^{BB,\sat} + C_{l_1}^{BB,t} + \tilde{C}_{l_1}^{BB} - C_{l_1}^W\right).
    \end{equation}
    
\subsection{Additional corrections from internal delensing}\label{appendix:calculation_of_eb_bias}
    
    We now consider the remaining couplings, represented by the diagrams of eqs.~\eqref{eqn:temp_x_obs_diags} and~\eqref{eqn:temp_x_temp_diags}, which are not included in the standard calculation and lead to biases if left uncorrected. 
    
    We start with the six-point function (eq.~\ref{eqn:temp_x_temp_diags}).
    The contribution from the connected six-point function (the first diagram on the right) is expected to be negligible compared to the other terms in the six-point function since it is one order higher in 
    $C_l^{\phi\phi}$. For the other terms, we are guided by the argument above about the volume of wavevector space available given the implied couplings between the weights. For the remaining terms involving the connected four-point function,
    this suggests that, symbolically,
    \begin{equation}
    \corrIV \quad \gg \quad \corrVI \quad , \quad \corrIII \quad , \quad \corrI \,\, ,
    \label{eq:fourptbiggest}
    \end{equation}
    so we retain only the term on the left.
    The simulations of ref.~\cite{ref:namikawa_17} indicate that the first term makes the largest contribution of those on the right of eq.~\eqref{eq:fourptbiggest}, but it is still significantly smaller (on large scales) than that on the left.
    Retaining only the primary coupling, we find
    \begin{align}\label{eqn:corrIV}
    2\,\times\,\corrIV \quad \supset& \quad 2\,\wick{\langle \c1 E^{\text{obs}} \c1 E^{\text{obs}} \tilde{B}[\c2 E,\c3 \phi] \tilde{E}[\c2 E,\phi] \tilde{E}[\c4 E,\phi] \tilde{B}[\c4 E,\c3 \phi]\rangle}\nonumber\\
    \rightarrow & \quad 2\,(2\pi)^2 \delta^{(2)}(\bl_1+\bl_2)C_{l_1}^{W}D_{l_1} \, ,
    \end{align}
    where we have defined
    \begin{equation}
    D_{l} \equiv \frac{1}{C_{l}^{BB,\obs,\fid,\lat}} \int \measp W^2(\bl,\bl')\left[\We_{l'}\tilde{C}_{l'}^{EE}\right]\left[\Wp_{|\bl-\bl'|}A_{|\bl-\bl'|}^{EB}\right] \, .
    \end{equation}
    Note that the spectrum $D_l$ arises when one contracts the two observed $E$-modes that explicitly appear in the lensing template:
    \begin{equation}\label{eqn:avtemplensE}
    \langle B^{\temp}(\bl) \rangle_{E^{\obs,\lat}} = B^{\obs,\lat}(\bl) D_l \, .  
    \end{equation}
    For the remaining terms in eq.~\eqref{eqn:temp_x_temp_diags} that involve only the two-point functions, we expect, symbolically, that
    \begin{equation}
    \corrII  \quad \gg \quad \corrforgottenI \, .
    \end{equation}
    Retaining only the term on the left, we have
    \begin{equation}\label{eqn:corrII}
    \corrII  \quad \rightarrow\quad  (2\pi)^2 \delta^{(2)}(\bl_1+\bl_2)C^{BB,\obs,\lat}_{l_1} D_{l_1}^2 \quad.
    \end{equation}
    Note how this involves a contribution from the primordial $B$-mode power, if present, through $C^{BB,\obs,\lat}_{l_1}$.
    
    Finally, we return to the four-point function, eq.~\eqref{eqn:temp_x_obs_diags}. The term not included in the standard calculation is that involving only the two-point functions, whose contribution to the delensed power spectrum is
    \begin{equation}\label{eqn:corrtempxobsdisc}
    -\,2\,\times\, \corrtempxobsdisc \quad \rightarrow\quad 
    -2 D_{l_1}\langle B^{\obs,\lat}(\bl_1) B^{\obs,\sat}(\bl_2) \rangle \, .
    \end{equation}
    The term $D_{l_1} B^{\obs,\lat}(\bl_1)$ arises from contracting the observed $E$-modes in the template represented by the left-half of the diagram. Allowing for the case where the $B$-modes used in the lensing reconstruction are from the same survey as those used to measure the large-scale $B$-modes, with the noise cross-spectrum $N_l^X$, we have
    \begin{equation}\label{eqn:corrtempxobsdiscii}
     -2 D_{l_1}\langle B^{\obs,\lat}(\bl_1) B^{\obs,\sat}(\bl_2) \rangle
    = - 2\,(2\pi)^2 \delta^{(2)}(\bl_1+\bl_2)\left(\tilde{C}^{BB}_{l_1}+{C}^{BB,\mathrm{t}}_{l_1}+{N}^{X}_{l_1}\right) D_{l_1} \, .
    \end{equation}
    Note how this term also receives a contribution from the primordial $B$-mode power.
    
    The term~\eqref{eqn:corrtempxobsdiscii} differs from its equivalent in previous work. In ref.~\cite{ref:namikawa_nagata_14}, only the lensing contribution on the right is considered (see the third term on the right of their eq.~A16), while in ref.~\cite{ref:teng_11} only the tensor and noise contributions are present (see their eqs. 7 and 8).
    The total correction to the standard delensed power involving the tensor $B$-mode power is $D_l(D_l-2){C}^{BB,\mathrm{t}}_{l}$, which is negative since $0 < D_l < 1$ (see the discussion after eq.~\ref{eqn:D_l} in the main text) and so there is a suppression of the primordial power~\cite{ref:teng_11}.

\begin{figure}
        \centering
        \includegraphics[width=0.7\textwidth]{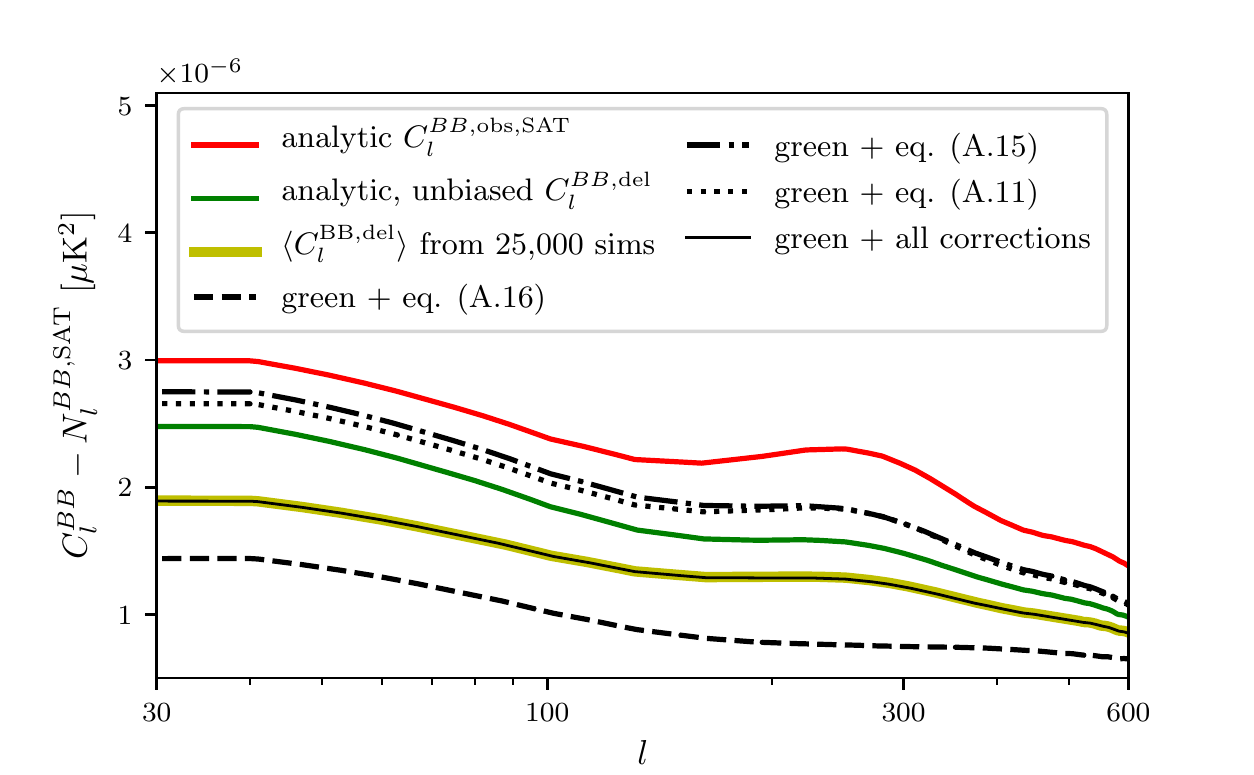}
        \caption{Break-down of contributions to the model, biased, delensed power spectrum of eq.~\eqref{eqn:bias_two_telescopes} in the case where $r_{\text{input}}=0.01$. The experimental set-up is a LAT with polarisation noise $\Delta_{\text{P}}=6\sqrt{2}\,\mu \mathrm{K\,arcmin}$ and beam size $\theta_{\mathrm{FWHM}}=1.5\, \mathrm{arcmin}$ and a SAT with $\Delta_{\text{P}}=2\sqrt{2}\,\mu \mathrm{K\,arcmin}$ and $\theta_{\mathrm{FWHM}}=17\, \mathrm{arcmin}$. The average delensed power from 25\,000 simulations is also shown.}
        \label{fig:all_bias_terms}
    \end{figure}

    Putting all the relevant terms together, the biased delensed spectrum can be modelled as
    \begin{align}\label{eqn:bias_two_telescopes}
        C_l^{BB,\text{del}} & = C_l^{BB,\obs,\sat} - C_l^W + C_l^{BB,\obs,\lat}D_l^2 - 2D_l(\tilde{C}_l^{BB} + C_l^{BB,t}+N_l^X -C_l^W) \nonumber\\
    & = (C_l^{BB,\text{res}} + C_l^{BB,\mathrm{t}})(D_l-1)^2 + D_l^2C_l^W + N_l^{BB,\sat} +  N_l^{BB,\lat}D_l^2 - 2D_l N_l^X \, .
    \end{align}
    The contributions from the various terms are shown in figure~\ref{fig:all_bias_terms}, along with the total delensed power measured from simulations. The model for the total predicted power agrees very well with the simulation results.

    The bias, i.e., the terms in eq.~\eqref{eqn:bias_two_telescopes} involving $D_l$, can be avoided by excluding from the lensing reconstruction any $B$-modes that overlap in scale with the $B$-modes we wish to delens~\cite{ref:teng_11,ref:namikawa_nagata_14}. This easily follows from noting that all significant bias terms arise from contracting the pair of observed $E$-modes in at least one lensing template, that is they involve $\langle B^{\temp}(\bl) \rangle_{E^{\obs,\lat}}$ when attempting to delens $B$-modes at wavevector $\bl$. The result of this contraction is proportional to the observed $B$-mode at $\bl$, $B^{\obs,\lat}(\bl)$, used in the lensing reconstruction (see eq.~\ref{eqn:avtemplensE}). If such modes are excluded from the reconstruction, the bias necessarily vanishes.
    
\begin{figure}
        \centering
        \includegraphics[width=0.7\textwidth]{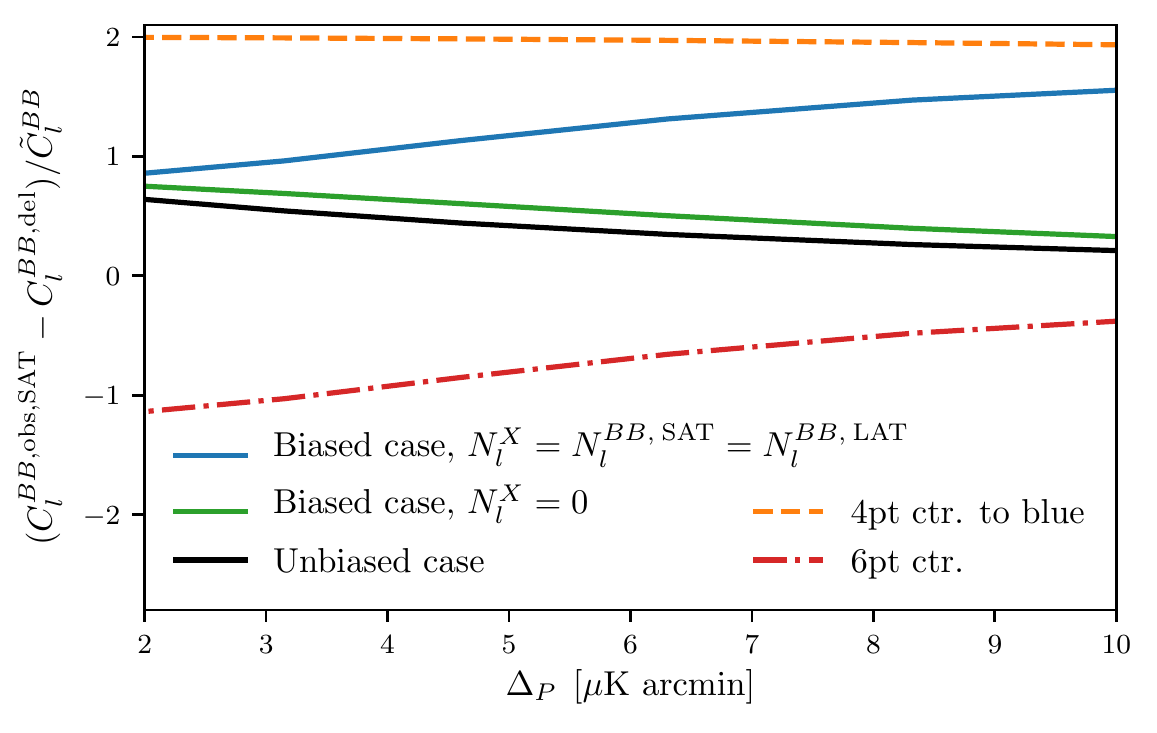}
        \caption{Break down of contributions to the difference between the power spectrum of the $B$-modes observed with the SAT and the power spectrum of the delensed $B$-modes, as a fraction of the original $B$-mode lensing power. This apparent delensing efficiency is shown as function of the polarisation white-noise level, for vanishing $r_{\rm input}$, keeping other specifications unchanged with respect to our baseline configuration in figure~\ref{fig:reconstruction_noise_levels}. In the case of independent surveys, the same noise level is assumed in each ($N_l^X=0$ and $N_l^{BB,\sat}=N_l^{BB,\lat}$). The internal delensing bias is larger for a single survey (blue) than for independent \sat~and \lat\ noise (green), in both cases artificially inflating the apparent delensing efficiency compared to the unbiased case (black).  The orange line shows the contribution to the single-survey case (i.e., the blue line) from the cross-correlation between the template and the observed $B$-modes, which is almost exactly equal to the full lensing power  (see eq.~\ref{eq:crossspec}). The red line shows the contribution from the auto-power spectrum of the template.} 
        \label{fig:bias_compplot}
    \end{figure}

    In the case of a single survey, the correlated noise between the $B$-modes used in the lensing reconstruction and the $B$-modes to delens sources a larger bias on the delensed spectrum than in the case of independent surveys (although, as explained in section~\ref{sec:biases}, the signal-to-noise on primordial $B$-mode power is also greater in the former configuration), and acts to emulate an apparent, but spurious, delensing efficiency much greater than expected. This is particularly clear from the cross-spectrum of the template with the observed $B$-modes. Combining eq.~\eqref{eqn:corrtempxobsdiscii} -- after identifying $D_l$ with $C_l^{BB, \mathrm{res}}/C_{l}^{BB,\obs,\fid,\lat}$ in the limit where $E$-mode noise can be neglected in the template (see
    section~\ref{sec:biases}) -- with the unbiased result eq.~\eqref{eqn:contrib_to_std_from_tempxobsconn}, we find that this cross-spectrum produces the entire lensing power:
\begin{equation}\label{eq:crossspec}
    \langle B^{\obs,\,\sat}(\bl_1) B^{\rm templ,\,\sat}(\bl_2)\rangle \approx (2\pi)^2 \delta^{(2)}(\bl_1 + \bl_2)\tilde C^{BB}_{l_1}\quad	 \text{(for $N_l^{X} = N_l^{BB, \lat} = N_l^{BB, \sat}$)}
\end{equation}
irrespective of the actual fidelity of the lensing reconstruction. In figure~\ref{fig:bias_compplot}, we quantify further this apparent delensing. We show the difference of the power spectrum of the observed \sat\ $B$-modes and the power spectrum of their delensed counterparts as a fraction of the $B$-mode lensing power as the instrument noise level is varied. This apparent delensing efficiency is shown averaged over degree-scale multipoles, for $r_{\rm input} = 0$, keeping otherwise the same simulation and reconstruction parameters as in figure~\ref{fig:reconstruction_noise_levels}. It is shown for the single-survey case in blue, while the black and green curves show the unbiased case ($C_l^{W} / \tilde{C}^{BB}_l$) and biased case with $N_l^X =0$, respectively, for comparison. The single-survey apparent efficiency is broken down into the contribution from the four-point function of the observed fields (i.e., twice the cross-correlation between the template and the observed $B$-modes; shown in orange) and the six-point function (i.e., the auto-power spectrum of the template; red). The latter contribution does not depend on $N_l^X$ and so is the same whether the surveys are independent or not (provided $N_l^{BB,\sat} = N_l^{BB,\lat}$), but the contribution of the cross-spectrum is boosted in the single-survey case. In this case, at low noise levels both cross- and auto-spectra are conspicuously close to $\tilde C^{BB}_l$, resulting in an apparent delensing efficiency close to 100\,\%. For higher noise levels, the apparent delensing efficiency is even larger, exceeding 100\,\%.

\section{Simulations}\label{appendix:sims}
    We simulate observations of the CMB sky using the publicly-available code \texttt{QuickLens}\footnote{\texttt{https://github.com/dhanson/quicklens}, though an amended and extended version can be found at \texttt{https://github.com/abaleato/Quicklens-with-fixes}.} for a fiducial cosmology best fitting the Planck+WP+highL data of ref.~\citep{ref:planck_14_params}. Inspired by the experimental configurations of the upcoming Simons Observatory, we simulate a reconstruction-oriented large-aperture-telescope (LAT) survey with noise levels $\Delta_{P}=6\sqrt{2}\, \mu\text{K\,arcmin} = \sqrt{2}\,\Delta_{T}$ and beam FWHM of $ \theta_{\mathrm{FWHM}} =1.5\,\mathrm{arcmin} $, together with a small-aperture-telescope (SAT) survey with $\Delta_{P}=2\sqrt{2}\, \mu\text{K\,arcmin} = \sqrt{2}\,\Delta_{T}$ and $ \theta_{\mathrm{FWHM}} =17\,\mathrm{arcmin} $. The simulations we generate are on the flat sky with 1024 pixels per side with an inter-pixel separation of $2\,\mathrm{arcmin}$, covering approximately $2.8\,\%$ of the sky. Our simulations have periodic boundary conditions and are free of foregrounds. Note that we only simulate the part of the LAT survey that overlaps with the SAT survey.
    
    The procedure for obtaining lensed CMB maps is as follows: first, we generate unlensed $ T,E,B $ and $ \phi $ fields in harmonic space by drawing their Fourier coefficients from zero-mean Gaussian distributions with variance at each (interpolated) angular scale given by the theory power spectra (obtained from \texttt{CAMB}~\citep{ref:lewis_challinor_lasenby_99}). At this stage, we choose a pixelisation, which in turn sets the maximum frequency we can adequately sample by the Nyquist--Shannon sampling theorem. Then, the unlensed $ T,E $ and $ B $ fields are rotated into $ T,Q $ and $ U $ and remapped according to the deflection field $ \bm{d}(\hat{\bm{n}})=\bm{\nabla}\phi $ as 
    \begin{equation}
    \tilde{T}=T[\hat{\bm{n}} + \bm{d}(\hat{\bm{n}})],
    \end{equation}
    and analogously for $ Q $ and $ U $, using a bivariate spline interpolation over a rectangular mesh. In order to mimic the effect of observations, we convolve the lensed fields with the transfer function for an experimental beam that is assumed to be symmetric and Gaussian with the required FWHM.
    As a final step, we add uncorrelated, Gaussian-distributed experimental noise at the map level.

\bibliographystyle{JHEP}
\bibliography{paper}

\newcommand{\noop}[1]{}

\providecommand{\href}[2]{#2}\begingroup\raggedright\begin{thebibliography}{10}

\bibitem{Polnarev:1985}
A.~G. {Polnarev}, \emph{{Polarization and Anisotropy Induced in the Microwave
  Background by Cosmological Gravitational Waves}}, {\emph{\sovast} {\bfseries
  29} (1985) 607}.

\bibitem{Kamionkowski:1996zd}
M.~Kamionkowski, A.~Kosowsky and A.~Stebbins, \emph{{A Probe of primordial
  gravity waves and vorticity}},
  \href{https://doi.org/10.1103/PhysRevLett.78.2058}{\emph{Phys. Rev. Lett.}
  {\bfseries 78} (1997) 2058}
  [\href{https://arxiv.org/abs/astro-ph/9609132}{{\ttfamily
  astro-ph/9609132}}].

\bibitem{Seljak:1996gy}
U.~Seljak and M.~Zaldarriaga, \emph{{Signature of gravity waves in polarization
  of the microwave background}},
  \href{https://doi.org/10.1103/PhysRevLett.78.2054}{\emph{Phys. Rev. Lett.}
  {\bfseries 78} (1997) 2054}
  [\href{https://arxiv.org/abs/astro-ph/9609169}{{\ttfamily
  astro-ph/9609169}}].

\bibitem{ref:bicep2_18}
{BICEP2 Collaboration}, {Keck Array Collaboration}, P.~A.~R. {Ade}, Z.~{Ahmed},
  R.~W. {Aikin}, K.~D. {Alexand er} et~al., \emph{{Constraints on Primordial
  Gravitational Waves Using Planck, WMAP, and New BICEP2/Keck Observations
  through the 2015 Season}},
  \href{https://doi.org/10.1103/PhysRevLett.121.221301}{\emph{\prl} {\bfseries
  121} (2018) 221301} [\href{https://arxiv.org/abs/1810.05216}{{\ttfamily
  1810.05216}}].

\bibitem{Zaldarriaga:1998ar}
M.~Zaldarriaga and U.~Seljak, \emph{{Gravitational lensing effect on cosmic
  microwave background polarization}},
  \href{https://doi.org/10.1103/PhysRevD.58.023003}{\emph{Phys. Rev.}
  {\bfseries D58} (1998) 023003}
  [\href{https://arxiv.org/abs/astro-ph/9803150}{{\ttfamily
  astro-ph/9803150}}].

\bibitem{ref:hanson_13}
{SPT Collaboration}, \emph{{Detection of B-Mode Polarization in the Cosmic
  Microwave Background with Data from the South Pole Telescope}},
  \href{https://doi.org/10.1103/PhysRevLett.111.141301}{\emph{\prl} {\bfseries
  111} (2013) 141301}.

\bibitem{ref:carron_17}
J.~{Carron}, A.~{Lewis} and A.~{Challinor}, \emph{{Internal delensing of Planck
  CMB temperature and polarization}},
  \href{https://doi.org/10.1088/1475-7516/2017/05/035}{\emph{Journal of
  Cosmology and Astro-Particle Physics} {\bfseries 2017} (2017) 035}.

\bibitem{ref:spt_17}
{SPT Collaboration}, \emph{{CMB Polarization B-mode Delensing with SPTpol and
  Herschel}}, \href{https://doi.org/10.3847/1538-4357/aa82bb}{\emph{\apj}
  {\bfseries 846} (2017) 45}.

\bibitem{Planck2018:lensing}
{Planck Collaboration}, \emph{{Planck 2018 results. VIII. Gravitational
  lensing}}, {\emph{arXiv e-prints} (2018) arXiv:1807.06210}
  [\href{https://arxiv.org/abs/1807.06210}{{\ttfamily 1807.06210}}].

\bibitem{ref:polarbear_delensing_19}
S.~{Adachi}, M.~A.~O. {Aguilar Fa{\'u}ndez}, Y.~{Akiba}, A.~{Ali}, K.~{Arnold},
  C.~{Baccigalupi} et~al., \emph{{Internal delensing of cosmic microwave
  background polarization B-modes with the POLARBEAR experiment}}, {\emph{arXiv
  e-prints} (2019) arXiv:1909.13832}
  [\href{https://arxiv.org/abs/1909.13832}{{\ttfamily 1909.13832}}].

\bibitem{ref:han_20}
D.~{Han}, N.~{Sehgal}, A.~{MacInnis}, A.~{van Engelen}, B.~D. {Sherwin}, M.~S.
  {Madhavacheril} et~al., \emph{{The Atacama Cosmology Telescope: delensed
  power spectra and parameters}},
  \href{https://doi.org/10.1088/1475-7516/2021/01/031}{\emph{\jcap} {\bfseries
  2021} (2021) 031} [\href{https://arxiv.org/abs/2007.14405}{{\ttfamily
  2007.14405}}].

\bibitem{ref:bicep_delensing}
{BICEP/Keck}, {SPTpol Collaborations}, {:}, P.~A.~R. {Ade}, Z.~{Ahmed},
  M.~{Amiri} et~al., \emph{{A Demonstration of Improved Constraints on
  Primordial Gravitational Waves with Delensing}}, {\emph{arXiv e-prints}
  (2020) arXiv:2011.08163} [\href{https://arxiv.org/abs/2011.08163}{{\ttfamily
  2011.08163}}].

\bibitem{ref:template_B_paper_in_prep}
A.~{Baleato Lizancos}, A.~{Challinor} and J.~{Carron}, \emph{{Limitations of
  CMB B -mode template delensing}},
  \href{https://doi.org/10.1103/PhysRevD.103.023518}{\emph{\prd} {\bfseries
  103} (2021) 023518} [\href{https://arxiv.org/abs/2010.14286}{{\ttfamily
  2010.14286}}].

\bibitem{ref:planck_template}
{Planck Collaboration}, P.~A.~R. {Ade}, N.~{Aghanim}, M.~{Ashdown},
  J.~{Aumont}, C.~{Baccigalupi} et~al., \emph{{Planck intermediate results.
  XLI. A map of lensing-induced B-modes}},
  \href{https://doi.org/10.1051/0004-6361/201527932}{\emph{\aap} {\bfseries
  596} (2016) A102} [\href{https://arxiv.org/abs/1512.02882}{{\ttfamily
  1512.02882}}].

\bibitem{ref:okamoto_hu_2003}
T.~{Okamoto} and W.~{Hu}, \emph{{Cosmic microwave background lensing
  reconstruction on the full sky}},
  \href{https://doi.org/10.1103/PhysRevD.67.083002}{\emph{\prd} {\bfseries 67}
  (2003) 083002}.

\bibitem{ref:hirata_03_polarization}
C.~M. {Hirata} and U.~{Seljak}, \emph{{Reconstruction of lensing from the
  cosmic microwave background polarization}},
  \href{https://doi.org/10.1103/PhysRevD.68.083002}{\emph{\prd} {\bfseries 68}
  (2003) 083002}.

\bibitem{ref:carron_17_maximum}
J.~{Carron} and A.~{Lewis}, \emph{{Maximum a posteriori CMB lensing
  reconstruction}},
  \href{https://doi.org/10.1103/PhysRevD.96.063510}{\emph{\prd} {\bfseries 96}
  (2017) 063510}.

\bibitem{ref:millea_17}
M.~{Millea}, E.~{Anderes} and B.~D. {Wandelt}, \emph{{Bayesian delensing of CMB
  temperature and polarization}}, {\emph{ArXiv e-prints} (2017)
  arXiv:1708.06753} [\href{https://arxiv.org/abs/1708.06753}{{\ttfamily
  1708.06753}}].

\bibitem{ref:teng_11}
W.-H. {Teng}, C.-L. {Kuo} and J.-H. {Proty Wu}, \emph{{Cosmic Microwave
  Background Delensing Revisited: Residual Biases and a Simple Fix}},
  {\emph{ArXiv e-prints} (2011) arXiv:1102.5729}
  [\href{https://arxiv.org/abs/1102.5729}{{\ttfamily 1102.5729}}].

\bibitem{ref:so_science_paper}
{The Simons Observatory Collaboration}, \emph{{The Simons Observatory: science
  goals and forecasts}},
  \href{https://doi.org/10.1088/1475-7516/2019/02/056}{\emph{\jcap} {\bfseries
  2019} (2019) 056} [\href{https://arxiv.org/abs/1808.07445}{{\ttfamily
  1808.07445}}].

\bibitem{ref:smith_12_external}
K.~M. {Smith}, D.~{Hanson}, M.~{LoVerde}, C.~M. {Hirata} and O.~{Zahn},
  \emph{{Delensing CMB polarization with external datasets}},
  \href{https://doi.org/10.1088/1475-7516/2012/06/014}{\emph{\jcap} {\bfseries
  2012} (2012) 014}.

\bibitem{ref:sherwin_15}
B.~D. {Sherwin} and M.~{Schmittfull}, \emph{{Delensing the CMB with the cosmic
  infrared background}},
  \href{https://doi.org/10.1103/PhysRevD.92.043005}{\emph{\prd} {\bfseries 92}
  (2015) 043005}.

\bibitem{ref:larsen_16}
P.~{Larsen}, A.~{Challinor}, B.~D. {Sherwin} and D.~{Mak}, \emph{{Demonstration
  of Cosmic Microwave Background Delensing Using the Cosmic Infrared
  Background}},
  \href{https://doi.org/10.1103/PhysRevLett.117.151102}{\emph{\prl} {\bfseries
  117} (2016) 151102}.

\bibitem{Schmittfull:2017ffw}
M.~Schmittfull and U.~Seljak, \emph{{Parameter constraints from
  cross-correlation of CMB lensing with galaxy clustering}},
  \href{https://doi.org/10.1103/PhysRevD.97.123540}{\emph{Phys. Rev. D}
  {\bfseries 97} (2018) 123540}
  [\href{https://arxiv.org/abs/1710.09465}{{\ttfamily 1710.09465}}].

\bibitem{ref:s4}
{CMB-S4 collaboration}, \emph{{CMB-S4 Science Book, First Edition}},
  {\emph{ArXiv e-prints} (2016) arXiv:1610.02743}
  [\href{https://arxiv.org/abs/1610.02743}{{\ttfamily 1610.02743}}].

\bibitem{ref:hirata_03_temperature}
C.~M. {Hirata} and U.~{Seljak}, \emph{{Analyzing weak lensing of the cosmic
  microwave background using the likelihood function}},
  \href{https://doi.org/10.1103/PhysRevD.67.043001}{\emph{\prd} {\bfseries 67}
  (2003) 043001}.

\bibitem{ref:core_lensing}
A.~{Challinor}, R.~{Allison}, J.~{Carron}, J.~{Errard}, S.~{Feeney},
  T.~{Kitching} et~al., \emph{{Exploring cosmic origins with CORE:
  Gravitational lensing of the CMB}},
  \href{https://doi.org/10.1088/1475-7516/2018/04/018}{\emph{\jcap} {\bfseries
  2018} (2018) 018} [\href{https://arxiv.org/abs/1707.02259}{{\ttfamily
  1707.02259}}].

\bibitem{ref:hu_okamoto_2002}
W.~{Hu} and T.~{Okamoto}, \emph{{Mass Reconstruction with Cosmic Microwave
  Background Polarization}}, \href{https://doi.org/10.1086/341110}{\emph{\apj}
  {\bfseries 574} (2002) 566}.

\bibitem{ref:spt3g_14}
{SPT-3G Collaboration}, \emph{{SPT-3G: a next-generation cosmic microwave
  background polarization experiment on the South Pole telescope}},  in
  \emph{Millimeter, Submillimeter, and Far-Infrared Detectors and
  Instrumentation for Astronomy VII}, W.~S. {Holland} and J.~{Zmuidzinas},
  eds., vol.~9153 of \emph{Society of Photo-Optical Instrumentation Engineers
  (SPIE) Conference Series}, p.~91531P, July, 2014,
  \href{https://doi.org/10.1117/12.2057305}{DOI}
  [\href{https://arxiv.org/abs/1407.2973}{{\ttfamily 1407.2973}}].

\bibitem{ref:hanson_11}
D.~{Hanson}, A.~{Challinor}, G.~{Efstathiou} and P.~{Bielewicz}, \emph{{CMB
  temperature lensing power reconstruction}},
  \href{https://doi.org/10.1103/PhysRevD.83.043005}{\emph{\prd} {\bfseries 83}
  (2011) 043005}.

\bibitem{ref:lewis_11}
A.~{Lewis}, A.~{Challinor} and D.~{Hanson}, \emph{{The shape of the CMB lensing
  bispectrum}},
  \href{https://doi.org/10.1088/1475-7516/2011/03/018}{\emph{\jcap} {\bfseries
  2011} (2011) 018} [\href{https://arxiv.org/abs/1101.2234}{{\ttfamily
  1101.2234}}].

\bibitem{ref:challinor_2005}
A.~{Challinor} and A.~{Lewis}, \emph{{Lensed CMB power spectra from all-sky
  correlation functions}},
  \href{https://doi.org/10.1103/PhysRevD.71.103010}{\emph{\prd} {\bfseries 71}
  (2005) 103010}.

\bibitem{ref:cooray_kesden_03}
A.~{Cooray} and M.~{Kesden}, \emph{{Weak lensing of the CMB: extraction of
  lensing information from the trispectrum}},
  \href{https://doi.org/10.1016/S1384-1076(02)00225-7}{\emph{\na} {\bfseries 8}
  (2003) 231} [\href{https://arxiv.org/abs/astro-ph/0204068}{{\ttfamily
  astro-ph/0204068}}].

\bibitem{ref:namikawa_15}
T.~{Namikawa} and R.~{Nagata}, \emph{{Non-Gaussian structure of B-mode
  polarization after delensing}},
  \href{https://doi.org/10.1088/1475-7516/2015/10/004}{\emph{\jcap} {\bfseries
  2015} (2015) 004}.

\bibitem{ref:hamimeche_2008}
S.~{Hamimeche} and A.~{Lewis}, \emph{{Likelihood analysis of CMB temperature
  and polarization power spectra}},
  \href{https://doi.org/10.1103/PhysRevD.77.103013}{\emph{\prd} {\bfseries 77}
  (2008) 103013}.

\bibitem{ref:namikawa_nagata_14}
T.~{Namikawa} and R.~{Nagata}, \emph{{Lensing reconstruction from a patchwork
  of polarization maps}},
  \href{https://doi.org/10.1088/1475-7516/2014/09/009}{\emph{\jcap} {\bfseries
  2014} (2014) 009}.

\bibitem{ref:namikawa_17}
T.~{Namikawa}, \emph{{CMB internal delensing with general optimal estimator for
  higher-order correlations}},
  \href{https://doi.org/10.1103/PhysRevD.95.103514}{\emph{\prd} {\bfseries 95}
  (2017) 103514}.

\bibitem{ref:sehgal_17}
N.~{Sehgal}, M.~S. {Madhavacheril}, B.~{Sherwin} and A.~{van Engelen},
  \emph{{Internal delensing of cosmic microwave background acoustic peaks}},
  \href{https://doi.org/10.1103/PhysRevD.95.103512}{\emph{\prd} {\bfseries 95}
  (2017) 103512}.

\bibitem{ref:smith_2004}
K.~M. {Smith}, W.~{Hu} and M.~{Kaplinghat}, \emph{{Weak lensing of the CMB:
  Sampling errors on B modes}},
  \href{https://doi.org/10.1103/PhysRevD.70.043002}{\emph{\prd} {\bfseries 70}
  (2004) 043002}.

\bibitem{ref:smith_challinor_2006}
S.~{Smith}, A.~{Challinor} and G.~{Rocha}, \emph{{What can be learned from the
  lensed cosmic microwave background B-mode polarization power spectrum?}},
  \href{https://doi.org/10.1103/PhysRevD.73.023517}{\emph{\prd} {\bfseries 73}
  (2006) 023517}.

\bibitem{ref:benoit_levy_12}
A.~{Benoit-L{\'e}vy}, K.~M. {Smith} and W.~{Hu}, \emph{{Non-Gaussian structure
  of the lensed CMB power spectra covariance matrix}},
  \href{https://doi.org/10.1103/PhysRevD.86.123008}{\emph{\prd} {\bfseries 86}
  (2012) 123008}.

\bibitem{ref:peloton_17}
J.~{Peloton}, M.~{Schmittfull}, A.~{Lewis}, J.~{Carron} and O.~{Zahn},
  \emph{{Full covariance of CMB and lensing reconstruction power spectra}},
  \href{https://doi.org/10.1103/PhysRevD.95.043508}{\emph{\prd} {\bfseries 95}
  (2017) 043508}.

\bibitem{ref:namikawa_lss}
T.~{Namikawa} and R.~{Takahashi}, \emph{{Impact of nonlinear growth of the
  large-scale structure on CMB B -mode delensing}},
  \href{https://doi.org/10.1103/PhysRevD.99.023530}{\emph{\prd} {\bfseries 99}
  (2019) 023530} [\href{https://arxiv.org/abs/1810.03346}{{\ttfamily
  1810.03346}}].

\bibitem{ref:cmb_likelihoods_review}
M.~{Gerbino}, M.~{Lattanzi}, M.~{Migliaccio}, L.~{Pagano}, L.~{Salvati},
  L.~{Colombo} et~al., \emph{{Likelihood methods for CMB experiments}},
  {\emph{arXiv e-prints} (2019) arXiv:1909.09375}
  [\href{https://arxiv.org/abs/1909.09375}{{\ttfamily 1909.09375}}].

\bibitem{ref:hu_2001_trispectrum}
W.~{Hu}, \emph{{Angular trispectrum of the cosmic microwave background}},
  \href{https://doi.org/10.1103/PhysRevD.64.083005}{\emph{\prd} {\bfseries 64}
  (2001) 083005}.

\bibitem{ref:kesden_cooray_kamionkowski_03}
M.~{Kesden}, A.~{Cooray} and M.~{Kamionkowski}, \emph{{Lensing reconstruction
  with CMB temperature and polarization}},
  \href{https://doi.org/10.1103/PhysRevD.67.123507}{\emph{\prd} {\bfseries 67}
  (2003) 123507} [\href{https://arxiv.org/abs/astro-ph/0302536}{{\ttfamily
  astro-ph/0302536}}].

\bibitem{ref:planck_14_params}
{Planck Collaboration}, \emph{{Planck 2013 results. XVI. Cosmological
  parameters}}, \href{https://doi.org/10.1051/0004-6361/201321591}{\emph{\aap}
  {\bfseries 571} (2014) A16}
  [\href{https://arxiv.org/abs/1303.5076}{{\ttfamily 1303.5076}}].

\bibitem{ref:lewis_challinor_lasenby_99}
A.~{Lewis}, A.~{Challinor} and A.~{Lasenby}, \emph{{Efficient Computation of
  Cosmic Microwave Background Anisotropies in Closed Friedmann-Robertson-Walker
  Models}}, \href{https://doi.org/10.1086/309179}{\emph{\apj} {\bfseries 538}
  (2000) 473}.

\end{thebibliography}\endgroup
\end{document}